\documentclass[aps,prd,preprintnumbers,superscriptaddress,nofootinbib,twocolumn]{revtex4-1}

\usepackage[dvipdfmx]{graphicx}
\usepackage{here}
\usepackage{braket}
\usepackage{feynmf}
\usepackage{xcolor}
\usepackage{bm,latexsym,amsmath,amssymb,amsfonts,mathrsfs}
\usepackage[colorlinks=true,urlcolor=teal,citecolor=orange,linkcolor=violet]{hyperref}
\usepackage{slashed}
\usepackage{framed} 

\usepackage{float}
\usepackage[caption = false]{subfig}
\usepackage{graphicx}

\newcommand*{\ba}{\begin{eqnarray}}
\newcommand*{\ea}{\end{eqnarray}}
\newcommand*{\bb}{\begin{framed}}
\newcommand*{\eb}{\end{framed}}

\newcommand*{\mpl}{M_{\rm Pl}}
\newcommand{\simgt}{\lower.5ex\hbox{$\; \buildrel > \over \sim \;$}}
\newcommand{\simlt}{\lower.5ex\hbox{$\; \buildrel < \over \sim \;$}}

\newcommand*{\p}{\partial}

\newcommand*{\calC}{{\cal C}}
\newcommand*{\calD}{{\cal D}}
\newcommand*{\calE}{{\cal E}}

\newcommand*{\calK}{{\cal K}}
\newcommand*{\calL}{{\cal L}}
\newcommand*{\calM}{{\cal M}}

\newcommand*{\calQ}{{\cal Q}}

\newcommand*{\calU}{{\cal U}}

\def\({\biggl(}
\def\){\biggr)}
\def\[{\biggl[}
\def\]{\biggr]}

\newcommand{\nn}{\nonumber \\}

\begin{document}

\title{Spin-2 dark matter from anisotropic Universe in bigravity}

\author{Yusuke \sc{Manita}}
\email{manita@tap.scphys.kyoto-u.ac.jp}
\affiliation{Department of Physics, Kyoto University, Kyoto 606-8502, Japan}

\author{Katsuki \sc{Aoki}}
\affiliation{Center for Gravitational Physics and Quantum Information, Yukawa Institute for Theoretical Physics, Kyoto University, 606-8502, Kyoto, Japan}

\author{Tomohiro \sc{Fujita}}
\affiliation{Waseda Institute for Advanced Study, Shinjuku, Tokyo 169-8050, Japan}

\author{Shinji \sc{Mukohyama}}
\affiliation{Center for Gravitational Physics and Quantum Information, Yukawa Institute for Theoretical Physics, Kyoto University, 606-8502, Kyoto, Japan}
\affiliation{Kavli Institute for the Physics and Mathematics of the Universe (WPI), The University of Tokyo, 277-8583, Chiba, Japan}

\preprint{KUNS-2946, YITP-22-150, IPMU22-0063}

 \date{\today}

\begin{abstract}
Bigravity is one of the natural extensions of general relativity and contains an additional massive spin-2 field which can be a good candidate for dark matter. To discuss the production of spin-2 dark matter, we study fixed point solutions of the background equations for axisymmetric Bianchi type-I Universes in two bigravity theories without Boulware-Deser ghost, i.e., Hassan-Rosen bigravity and Minimal Theory of Bigravity. We investigate the local and global stability of the fixed points and classify them. Based on the general analysis, we propose a new scenario  where spin-2 dark matter is produced by the transition from an anisotropic fixed point solution to isotropic one. The produced spin-2 dark matter can account for all or a part of dark matter and can be directly detected by laser interferometers in the same way as gravitational waves.
\end{abstract}

\maketitle

\section{Introduction}
Dark matter is an unknown matter component that accounts for more than 20\% of the total energy density in the current Universe~\cite{Planck:2018vyg}. Its true nature is still unknown and has been actively explored from both theoretical and observational perspectives. For the theoretical side, plentiful dark matter models with a broad mass range have been proposed, and various detection methods for each model have been suggested~\cite{Lin:2019uvt}. For example, ultralight bosonic field is one of the candidates for dark matter. A scalar field candidate with the lightest mass scale around $\mathcal{O}(10^{-21})$ eV is called fuzzy dark matter, and it is expected to solve the small-scale problems such as the core-cusp problem. The QCD axion, which  was originally introduced to solve the strong CP problem~\cite{Peccei:1977hh}, is also a scalar-type dark matter candidate, especially well motivated in a very light mass range, $m \ll 1$eV.

Since typical intrinsic characteristics of bosonic particles are mass and spin, it is natural to consider ultralight dark matter with nonzero spin. One of such extensions is dark photon, which is vector-type ultralight dark matter. Unlike scalar-type ultralight dark matter, dark photons include helicity-one modes. Recently, the phenomenology of the dark photon has been investigated, for example, its production mechanisms~\cite{Dror:2018pdh,Bastero-Gil:2018uel,Ema:2019yrd,Nakai:2020cfw,Salehian:2020asa,Firouzjahi:2020whk}, superradiance~\cite{East:2017ovw,Cardoso:2018tly,East:2017mrj}, etc. 

Furthermore, the tensor-type dark matter model called {\it spin-2 dark matter} has been proposed~\cite{Aoki:2016zgp,Aoki:2017cnz,Babichev:2016bxi,Babichev:2016hir}. 
Some production mechanisms of spin-2 dark matter have been investigated so far.
For example, generation by primordial magnetic fields \cite{Aoki:2017cnz}, bubble collision in the preheating era \cite{Aoki:2016zgp}, and misalignment mechanism \cite{Marzola:2017lbt}. 

Ultralight dark matter is also interesting from an observational point of view. Some ultralight dark matter models are expected to give detectable signals to gravitational wave interferometers. For example, axion-like particles couple to electromagnetic fields can cause the birefringence of the laser beams~\cite{DeRocco:2018jwe,Obata:2018vvr,Nagano:2019rbw}. The ultralight dark photon can also generates detectable signals by accelerating the mirrors of the gravitational wave detectors when it couples to baryonic matters~\cite{LIGOScientificCollaborationVirgoCollaboration:2021eyz,Michimura:2020vxn,Miller:2020vsl,Morisaki:2020gui}. Spin-2 dark matter can also leave detectable signals in the gravitational wave detectors by changing the effective length of the arm in a similar way to usual gravitational waves~\cite{Armaleo:2020efr}. 

Spin-2 dark matter is  closely related to massive gravity since it has a nonzero mass and couples to the matter fields as a usual graviton. Massive gravity has a long history, beginning with the pioneering work of linear massive gravity by Fierz and Pauli in 1939~\cite{Fierz:1939ix}. This theory can be generalized to the nonlinear level, but Ref.~\cite{Boulware:1972yco} found that non-linear massive gravity suffers from a ghost instability, which is often called the Boulware-Deser ghost. In 2010, the first ghost-free nonlinear massive gravity (dRGT theory) was proposed~\cite{deRham:2010ik,deRham:2010kj}, and it possesses five degrees of freedom. Motivated by difficulties in the cosmology of massive gravity~\cite{DeFelice:2012mx}, some extensions of dRGT theory have been explored. For example, Minimal Theory of Massive Gravity achieves to reduce the number of degrees of freedom to only two by imposing constraints~\cite{DeFelice:2015hla}. More recent development is the extension of Lorentz-invariant massive gravity, called generalized massive gravity and projected massive gravity~\cite{DeRham:2014wnv,Gumrukcuoglu:2020utx}. They can describe cosmic expansion without the initial strong coupling problem~\cite{Kenna-Allison:2020egn,Manita:2021qun}.
Massive gravity with single graviton can be a candidate for the origin of the accelerated expansion of the Universe, but it is difficult to construct a viable model of spin-2 dark matter based on massive gravity satisfying the strong mass constraint $m \lesssim 1.2 \times 10^{-22}\,\text{eV}$ from the gravitational wave observation of BH-BH merger~\cite{LIGOScientific:2016lio}. 
\footnote{The graviton mass bound for single massive gravity is summarized in~\cite{deRham:2016nuf} (see also~\cite{DeFelice:2021trp}). Some of them are stronger than the GW constraint but are model dependent.}

On the other hand, spin-2 dark matter can be originated from bigravity, which is a gravity theory with two dynamical metrics. The first proposal of bigravity without Boulware-Deser ghost is called {\it Hassan-Rosen bigravity}~\cite{Hassan:2011zd}, and was accomplished by extending dRGT massive gravity~\cite{deRham:2010ik,deRham:2010kj}. The Hassan-Rosen bigravity has seven degrees of freedom because it can be regarded as a nonlinear theory in which massive and massless gravitons are interacting with each other. On the other hand, {\it Minimal Theory of Bigravity}~\cite{DeFelice:2020ecp} is a ghost-free bigravity with only four degrees of freedom, which is constructed by extending the Minimal Theory of Massive Gravity~\cite{DeFelice:2015hla}.

In dRGT theory, Ref.~\cite{Gumrukcuoglu:2012aa} found that the background equations in Bianchi type-I Universe possess a fixed point solution and discussed an anisotropic FLRW Universe, in which each of the physical and fiducial metrics is homogeneous and isotropic but they do not share the same rotational Killing vectors and thus the system as a whole breaks the isotropy. In this paper, by extending the previous work, we find a fixed point with relatively large anisotropy for both Hassan-Rosen bigravity and the Minimal Theory of Bigravity. Moreover, by using this fixed point, we discuss a new scenario to produce spin-2 dark matter from the large anisotropy in the early universe. Since in bigravity, the anisotropic perturbation of FLRW universe can be regarded as spin-2 dark matter~\cite{Maeda:2013bha,Aoki:2017cnz}, if the early universe is anisotropic, it may give an initial amplitude for the spin-2 dark matter.

This paper is organized as follows. In Sec.~\ref{sec:bigravity}, we introduce two bigravity theories without Boulware-Deser ghost, i.e., Hassan-Rosen bigravity and Minimal Theory of Bigravity. In Sec.~\ref{sec:Anisotropic_fixed_point}, we consider the Bianch type-I Universe as an example of the anisotropic Universe, and show that the background equations are the same for both bigravity theories. We then find the anisotropic fixed point solutions, in which each metric is homogeneous and isotropic but they do not share the same rotational Killing vectors. In Sec.~\ref{sec:stability}, we classify the fixed points by their local stability and investigate their global stability by drawing the phase portraits around them. In Sec.~\ref{sec:DM_production}, as an implication of the anisotropic fixed point in bigravity, we discuss the production of the spin-2 dark matter and its detectability by gravitational wave interferometers. Section~\ref{sec:discussion} is devoted to conclusions.

\section{Bigravity}
\label{sec:bigravity}
Bigravity is one of the extensions of general relativity that has two dynamical metrics $g_{\mu\nu}$ and $f_{\mu\nu}$ interacting with each other. The action of bigravity is given by
\begin{align}
    S_g &= \frac{1}{2\kappa_g^2} \int d^{4} x \sqrt{-g} R^{(g)}+\frac{1}{2\kappa_f^2} \int d^{4} x \sqrt{-f} R^{(f)}
    \nn
    &+\frac{m^2}{\kappa^2} \int d^{4} x \calL_{\rm int}[g_{\mu\nu},f_{\mu\nu}]\,,
    \label{eq:bigravity_action}
\end{align}
where $R^{(g)}$ and $R^{(f)}$ are the Ricci scalars for $g_{\mu\nu}$ and $f_{\mu\nu}$, respectively. The first and second terms are the Einstein-Hilbert terms of the $g$-sector and the $f$-sector with the gravitational constants $\kappa_g^2$ and $\kappa_f^2$. The third term represents interactions between $g$-metric and $f$-metric, $m$ denotes a mass parameter, and $\kappa^2$ is defined by $\kappa^2:=\kappa_g^2+\kappa_f^2$. For later convenience, we also introduce the ratio of the gravitational constants
\begin{align}
    \alpha := \frac{\kappa_g}{\kappa_f}\,.
\end{align}
The interaction term depends on the model. At least two bigravity models without Boulware-Deser ghost have been proposed so far: the Hassan-Rosen bigravity (HRBG) \cite{Hassan:2011zd} and the Minimal Theory of Bigravity (MTBG) \cite{DeFelice:2020ecp}. In this section, we will briefly review those bigravity theories.

\subsection{Hassan-Rosen bigravity}
HRBG is the ghost-free bigravity which is constructed by extending the dRGT massive gravity~\cite{Hassan:2011zd}. The interaction term for HRBG is given by
\begin{align}
    \calL_{\rm int} = \sqrt{-g}\sum_{n=0}^{4}b_n e_n({\calK}) = \sqrt{-f}\sum_{n=0}^{4}b_{4-n} e_n({\tilde{\calK}})\,,
    \label{eq:interaction}
\end{align}
with constant parameters $b_k$. Here $\calK^\mu{}_\nu$ is defined as the root of
\begin{align}
    \calK^{\mu}{}_{\alpha}\calK^{\alpha}{}_{\nu} = g^{\mu\alpha}f_{\alpha\nu}\,,
\end{align}
$\tilde{\calK}^\mu{}_\nu$ is its inverse satisfying 
\begin{align}
\tilde{\calK}^{\mu}{}_{\alpha}\calK^{\alpha}{}_{\nu} = \delta^{\mu}_{\nu} = \calK^{\mu}{}_{\alpha}\tilde{\calK}^{\alpha}{}_{\nu}\,,\quad \tilde{\calK}^{\mu}{}_{\alpha}\tilde{\calK}^{\alpha}{}_{\nu} = f^{\mu\alpha}g_{\alpha\nu}\,,
\end{align}
and $e_n(\calM)$ denote the elementary symmetric polynomials of degree $n$ in the matrix $\calM^{\mu}{}_{\nu}$,
\begin{align}
    e_{0}(\calM)& = 1
    \,,\\
    e_{1}(\calM) &= [\calM]
    \,,\\
    e_{2}(\calM) &= \frac{1}{2!} ([\calM]^2-[\calM^2])
    \,, \\
    e_{3}(\calM) &= \frac{1}{3!}([\calM]^3-3[\calM][\calM^2]+2[\calM^3])
    \,, \\
    e_{4}(\calM) &= \frac{1}{4!}([\calM]^4-6[\calM]^2[\calM^2]+8[\calM][\calM^3]
    \nn
    &+3[\calM^2]^2-6[\calM^4])\,,
\end{align}
where the square bracket is the trace symbol, $[\calM]=\calM^{\mu}{}_{\mu},~[\calM^2]=\calM^{\mu}{}_{\nu}\calM^{\nu}{}_{\mu}$ and so on. The interaction term  \eqref{eq:interaction} would be unique to avoid the Boulware-Deser ghost under the Poincar\`{e} invariance \cite{Hassan:2011zd}. 

HRBG possesses $2+5$ physical degrees of freedom, corresponding to the massless graviton and the massive graviton. The massless graviton has two degrees of freedom as in general relativity while the massive graviton has five degrees of freedom corresponding to the helicity modes 0, $\pm$1, and $\pm$2.

\subsection{Minimal theory of bigravity}

Although a massive spin-2 field has five degrees of freedom under the Lorenz invariance, the physical degrees of freedom can be reduced to only two by breaking the Lorentz symmetry. The resultant theory is known as {\it Minimal Theory of Massive Gravity}~\cite{DeFelice:2015hla} and MTBG is its bigravity extension~\cite{DeFelice:2020ecp}. Similarly to HRBG, MTBG possesses one massless graviton and one massive graviton but the massive state only has two tensorial degrees of freedom in MTBG.

To construct the action of MTBG, we first adopt the ADM decompositions for both metrics
\begin{align}
    g_{\mu\nu} dx^{\mu} dx^{\nu}
    &= -N_g^2 dt^2+\gamma^{g}_{ij}(N_{g}^idt+dx^i)(N_{g}^j dt+dx^j)
    \,,\\
    f_{\mu\nu} dx^{\mu} dx^{\nu} 
    &= -N_f^2 dt^2+\gamma^{f}_{ij}(N_{f}^idt+dx^i)(N_{f}^j dt+dx^j)
    \,,
\end{align}
where $N_g,~N_f$ are the lapse functions and $N_g^i,~N_f^i$ are the shift vectors, and $\gamma^g_{ij},~\gamma^f_{ij}$ are the induced metrics on the constant-time hypersurface, respectively. We define the covariant derivatives on the time constant hypersurface, $\calD^g_i,~\calD^f_i$, associated with the $g$- and $f$-induced metrics $\gamma^g_{ij},~\gamma^f_{ij}$. The extrinsic curvatures are then
\begin{align}
    K^{g}_{ij} = \frac{1}{2N_g} (\p_t\gamma^{g}_{ij}-\calD^{g}_i N^{g}_j-\calD^{g}_j N^{g}_i)\,,
    \\
    K^{f}_{ij} = \frac{1}{2N_f} (\p_t\gamma^{f}_{ij}-\calD^{f}_i N^{f}_j-\calD^{f}_j N^{f}_i)\,.
\end{align}

The interaction Lagrangian in MTBG is composed of the precursor part and the constraint part
\begin{align}
    \calL_{\rm int} &=  \calL_{\rm int,prec}[\gamma^{g}_{ij},\gamma^{f}_{ij},\gamma_{g}^{ij},\gamma_{f}^{ij},K^{g}_{ij},K^{f}_{ij}]
    \nn
    &+\calL_{\rm int,const}[\gamma^{g}_{ij},\gamma^{f}_{ij},\gamma_{g}^{ij},\gamma_{f}^{ij},K^{g}_{ij},K^{f}_{ij}]\,.
\end{align}
They are explicitly given by
\begin{align}
    &\calL_{\rm int,prec}[\gamma^{g}_{ij},\gamma^{f}_{ij},\gamma_{g}^{ij},\gamma_{f}^{ij},K^{g}_{ij},K^{f}_{ij}] 
    \nn
    &= -\frac{1}{2}\left(N_g \sqrt{\gamma^{g}}\sum_{n=0}^{3} b_{n} e_{n}(\mathfrak{K}) +N_f \sqrt{\gamma^{f}} \sum_{n=0}^{3} b_{4-n} e_{n}(\tilde{\mathfrak{K}})\right)\,,
    \\
    &\calL_{\rm int, const}[\gamma^{g}_{ij},\gamma^{f}_{ij},\gamma_{g}^{ij},\gamma_{f}^{ij},K^{g}_{ij},K^{f}_{ij}] 
    \nn
    &= -\frac{1}{2}\Bigg[ \sqrt{\gamma^{g}} \mathcal{U}^{i}{}_{j}\mathcal{D}^g_{i} \lambda^{j}-\beta \sqrt{\gamma^{f}} \tilde{\mathcal{U}}^{i}{}_{j} \calD^f_{i} \lambda^{j}
    \nn
    &\qquad+\left(\lambda+\gamma_g^{i j}
    \calD^g_i \calD^g_j\bar{\lambda}\right) \sqrt{\gamma^g} \mathcal{U}^{k}{}_{l}\gamma_g^{lm} K^{g}_{mk}
    \nn &\qquad
    -\left(\lambda-\gamma_f^{i j} \calD^f_i \calD^f_j \bar{\lambda}\right) \sqrt{\gamma^{f}} \tilde{\mathcal{U}}^{k}{}_{l} \gamma_f^{lm} K^{f}_{mk}
    \nn
    &\qquad+\frac{m_g^{2}\left(\lambda+\gamma_g^{i j} \calD^g_i \calD^g_j \bar{\lambda}\right)^{2}}{4 N_g} \sqrt{\gamma^g}\left([\mathcal{U}^2]-\frac{1}{2}[\calU]^2\right)
    \nn
    &\qquad+\frac{m_f^{2}\left(\lambda-\gamma_f^{i j} \calD^f_i \calD^f_j  \bar{\lambda}\right)^{2}}{4 N_f} \sqrt{\gamma^f}\left([\tilde{\mathcal{U}}^2]-\frac{1}{2}[\tilde{\mathcal{U}}]^2 \right)\Bigg]\,,
\end{align}
where $\lambda,~\lambda^i,~\bar{\lambda},~\bar{\lambda}^i$ are the Lagrange multipliers, $\beta$ is a constant parameter, and
\begin{align}
    m_g:=m\frac{\kappa_g}{\kappa}=\frac{\alpha m}{\sqrt{1+\alpha^2}}\,,
    \quad
    m_f:=m\frac{\kappa_f}{\kappa}=\frac{m}{\sqrt{1+\alpha^2}}\,.
\end{align}
The matrix $\mathfrak{K}^i{}_j$ and its inverse $\tilde{\mathfrak{K}}^i{}_j$ are the roots of
\begin{align}
    \mathfrak{K}^i{}_k \mathfrak{K}^k{}_j = \gamma_g^{ik}\gamma^f_{kj}\,,
    \quad
    \tilde{\mathfrak{K}}^i{}_k \tilde{\mathfrak{K}}^k{}_j = \gamma_f^{ik}\gamma^g_{kj}\,,
\end{align}
and $\calU^i{}_j,\tilde{\calU}^i{}_j$ are the derivatives of the symmetric polynomials
\begin{align}
    \calU^i{}_j &:= \frac{1}{2} \sum_{n=0}^{3} b_{n} \left(\frac{\partial e_n(\mathfrak{K})}{\partial\mathfrak{K}^{j}{}_{i}} + \gamma_g^{ik}\gamma^g_{jl} \frac{\partial e_n(\mathfrak{K})}{\partial\mathfrak{K}^{k}{}_{l}}\right)\,,
    \\
    \tilde{\calU}^i{}_j &:= \frac{1}{2} \sum_{n=0}^{3} b_{4-n} \left(\frac{\partial e_n(\tilde{\mathfrak{K}})}{\partial\tilde{\mathfrak{K}}^{j}{}_{i}} + \gamma_f^{ik}\gamma^f_{jl} \frac{\partial e_n(\tilde{\mathfrak{K}})}{\partial\tilde{\mathfrak{K}}^{k}{}_{l}}\right)\,.
\end{align}
The precursor part possesses a structure similar to the interaction term of HRBG while the constraint part is added to eliminate the scalar and vector modes of the massive graviton. MTBG is constructed in such a way that background equations for a homogeneous universe coincide with those in HRBG. In Ref.~\cite{DeFelice:2020ecp}, this was checked only for the FLRW case. In the next section, we will show that the background equations are identical also for the Bianchi type-I Universe.

\section{Anisotropic universe}
\label{sec:Anisotropic_fixed_point}
Let us investigate the homogeneous and anisotropic universe both in HRBG and MTBG. For simplicity, we study the axisymmetric Bianchi type-I universe in vacuum. We first show that the background equations are identical in HRBG and MTBG, meaning that our following analysis can be applied to both bigravity theories. Then, we describe the generic structure of the equations of motion. These equations have several fixed points which we discuss in this section. The stability of the fixed points will be studied in the next section.

\subsection{Equations of motion}
For both $g$- and $f$- metric, we take a metric ansatz as one of the homogeneous Universe, Bianchi type-I spacetime
\begin{align}
    &g_{\mu\nu}dx^{\mu} dx^{\nu}
    \nn
    &= -N_{g}^{2} d t^{2}+a_{g}^{2}\left[e^{4 \beta_{g}} d x^{2}+e^{-2 \beta_{g}}\left(d y^{2}+d z^{2}\right)\right]\,,
    \\
    &f_{\mu\nu}dx^{\mu} dx^{\nu} 
    \nn
    &= -N_{f}^{2} d t^{2}+a_{f}^{2}\left[e^{4 \beta_{f}} d x^{2}+e^{-2 \beta_{f}}\left(d y^{2}+d z^{2}\right)\right]\,,
\end{align}
where the lapse functions $\{N_g, N_f\}$, the scale factors $\{a_g, a_f\}$, and the anisotropies $\{\beta_f, \beta_f\}$ are functions of time $t$. The $g$- and $f$-Hubble expansion rates and the shears are defined by
\begin{align}
    H_{g}&:=\frac{\dot{a}_{g}}{a_{g} N_{g}}\,, 
    \quad 
    H_{f}:=\frac{\dot{a}_{f}}{a_{f} N_{f}}\,,
    \\
    \sigma_{g}&:=\frac{\dot{\beta}_{g}}{N_{g}}\,, 
    \quad 
    \sigma_{f}:=\frac{\dot{\beta}_{f}}{N_{f}}\,.
\end{align}
For simplicity, we study vacuum solutions in the following.

The mini-superspace action in HRBG is given by~\cite{Maeda:2013bha}
\begin{align}
    S_{\rm mHRBG} = \frac{V}{2\kappa_g^2} &\int dt a_g^3 N_g 
    \nn
    \times \Bigg\{ &-6(H_g^2-\sigma_g^2) -6\alpha^2  r \xi^4 (H_f^2-\sigma_f^2) 
    \nn
    &+ m_g^2 \Big[ b_0  + b_1 \xi (r+e^{-2\beta}+2e^{\beta}) 
    \nn
    &+b_2\xi^2\left[2e^{-\beta}+e^{2\beta}+r(e^{-2\beta}+2e^{\beta})\right] 
    \nn
    &+ b_3\xi^3\left[1+r(2e^{-\beta}+e^{2\beta})\right]+b_4 r \xi^4 \Big] \Bigg\}\,,
    \label{eq:MiniSuperAction_HRBG}
\end{align}
where
\begin{align}
    \xi:=\frac{a_{f}}{a_{g}}\,\quad
     r:=\frac{a_g N_f}{a_f N_g}
     \,,
    \quad
    \beta:=\beta_g-\beta_f\,,
    \label{eq:defxir}
\end{align}
and $V \equiv \int d^3x$ formally represents the comoving volume of the system. Varying the action with respect to $X=\{N_g, ~ N_f, ~ a_g, ~a_f,~\beta_g,~\beta_f\}$, we obtain the background equations in the form $\calE_X=0$:
\begin{align}
    \calE_{N_g} &:=3 (H_{g}^{2} -  \sigma_{g}^{2})-m_g^{2}\Big[b_{0}+b_{1}\left(e^{-2 \beta}+2 e^{\beta}\right) \xi
    \nn &
    +b_{2}\left(2 e^{-\beta}+e^{2 \beta}\right) \xi^{2}+b_{3} \xi^{3}\Big]\,, 
    \label{eq:Friedmanng}
    \\
    \calE_{N_f} &:= 3 (H_{f}^{2} - \sigma_{f}^{2})-m_f^2 \Big[b_{4}+b_{3}\left(2 e^{-\beta}+e^{2 \beta}\right) \xi^{-1}   
    \nn
    &+b_{2}\left(e^{-2 \beta}+2 e^{\beta}\right) \xi^{-2}+b_{1} \xi^{-3}\Big]\,,
    \label{eq:Friedmannf}
    \\
    \calE_{a_g} &:= \frac{2\dot{H}_g}{N_g}+3(H_g^2+\sigma_g^2)
    \nn
    &-\frac{m_g^2}{3}\big\{ 3b_0 +b_1 \xi \left(3 r+2 e^{-2 \beta }+4 e^{\beta}\right)
    \nn
    &+b_2  \xi^2 \left[2r \left(2 e^{\beta}+e^{-2 \beta}\right)+ \left(e^{2 \beta}+2e^{-\beta}\right)\right]
    \nn
    &+b_3 r  \left(e^{2 \beta}+2e^{-\beta}\right) \xi^3\big\}\,,
    \label{eq:dotHg}
    \\
    \calE_{a_f} &:= \frac{2\dot{H}_f}{N_f}+3(H_f^2+\sigma_f^2)
   -\frac{m_f^2}{3 r\xi^3}\big\{  b_1 \left(e^{-2 \beta}+2 e^{\beta}\right)
    \nn
    &+b_2  \xi \left[r \left(e^{-2 \beta }+2 e^{\beta }\right)+2  \left(e^{2 \beta }+2e^{-\beta }\right)\right]
    \nn
    &+b_3 \xi^2 \left[2 r  \left(e^{2 \beta }+2e^{-\beta }\right)+3\right]+12 b_4 r \xi ^3 \big\}\,,
    \label{eq:dotHf}
    \\
    \calE_{\beta_g} &:= \frac{1}{a_{g}^{3}} \frac{d}{d t}\left(a_{g}^{3} \sigma_{g}\right)+\kappa_g^2 \frac{\partial U}{\partial \beta}\,, 
    \label{eq:anisotropy}
    \\
    \calE_{\beta_f} &:= \frac{1}{a_{g}^{3}} \frac{d}{d t}\left(a_{f}^{3} \sigma_{f}\right)-\kappa_f^2 \frac{\partial U}{\partial \beta}\,,
    \label{eq:anisotropyf}
\end{align}
where we have defined the potential of the anisotropy as
\begin{align}
    U := \frac{m^{2}}{6 \kappa^{2}}\big[ &\xi\left(2 e^{\beta}+e^{-2 \beta}\right)\left(b_{1}N_g+b_{2} N_{f}\right)
    \nn
    &+\xi^{2}\left(e^{2 \beta}+2 e^{-\beta}\right)\left(b_{2}N_g+b_{3} N_{f}\right)\big]\,.
\end{align}
As shown in the Friedmann equations \eqref{eq:Friedmanng} and \eqref{eq:Friedmannf}, the cosmic expansion is driven by the anisotropic shears $\sigma_g, \sigma_f$ and the graviton mass term. By using $\calE_{a_g}=0$ and $\calE_{\beta_g}=0$, we can eliminate $\dot{H}_g, \dot{\sigma}_g$ from $\dot{\calE}_{N_g}=0$ and then obtain the constraint equation $\calC=0$ with
\begin{align}
\calC&:=
    H_{g}\left[3 b_{1}+2 b_{2} \xi\left(2 e^{\beta}+e^{-2 \beta}\right)+b_{3} \xi^{2}\left(e^{2 \beta}+2 e^{-\beta}\right)\right]
    \nn
    &-H_{f} \xi\left[3 b_{3} \xi^{2}+2 b_{2} \xi\left(e^{2 \beta}+2 e^{-\beta}\right)+b_{1}\left(2 e^{\beta}+e^{-2 \beta}\right)\right]
    \nn
    &-2 \xi\left(e^{-\beta}-e^{2 \beta}\right)\left[\sigma_{f}\left(b_{1} e^{-\beta}+b_{2} \xi\right)+\sigma_{g}\left(b_{2} e^{-\beta}+b_{3} \xi\right)\right]\,.
    \nn
    \label{eq:constraint}
\end{align}
The same constraint equation is obtained by using $\dot{\calE}_{N_f}=0,~\calE_{a_f}=0$, and $\calE_{\beta_f}=0$ instead.

The minisuperspace action of MTBG is composed of the precursor part $S^{\rm mMTBG}_{\rm pre}$ and the constraint part $S_{\rm mMTBG}^{\rm const}$, where the precursor part agrees with the minisuperspace action of HRBG \eqref{eq:MiniSuperAction_HRBG} in the Bianchi type-I universe. 
The spatial homogeneity concludes that the spatial derivatives vanish and then the minisuperspace action does not depend on $\bar{\lambda}$ and $\lambda^i$. Then, the contribution of the constraint part to the mini-superspace action is given by a functional of $X=\{N_g, ~ N_f, ~ a_g, ~a_f,~\beta_g,~\beta_f\}$ and $\lambda(t)$:
\begin{align}
S_{\rm mMTBG}^{\rm const} = \frac{m^2V}{2\kappa^2} \int dt a_g^3 \left[ -\lambda \calC[X] + \frac{1}{2}\lambda^2 \calD[X] \right]
\,,
 \label{eq:action_const}
\end{align}
where $\calC$ is given in \eqref{eq:constraint} and $\calD$ is
\begin{align}
\calD&:=
\frac{m_f^2 e^{-4 \beta} }{4 N_g r \xi^2} \left(b_1+2 b_2 e^{\beta} \xi+b_3 e^{2 \beta} \xi^2\right)
\nn
&\times \Big[b_1 \left(3 \alpha ^2 r e^{4 \beta} \xi^2+4 e^{3 \beta }-1\right) 
    \nn
    &+2 b_2 e^{\beta} \xi \left[\alpha ^2 r e^{\beta} \left(e^{3 \beta}+2\right) \xi^2+2 e^{3 \beta}+1\right]
    \nn
    &+b_3 e^{2 \beta} \xi^2 \left[3-\alpha ^2 r e^{\beta} \left(e^{3 \beta}-4\right) \xi^2\right]\Big]
    \,.
\end{align}
The equations of motion are obtained by the variations of the total mini-superspace action $S_{\rm mMTBG}=S_{\rm mMTBG}^{\rm prec}+S_{\rm mMTBG}^{\rm const}$ with respect to $X$ and $\lambda$. Since the precursor part is identical to the minisuperspace action of HRBG, the equations of motion for $X=\{N_g, ~ N_f, ~ a_g, ~a_f,~\beta_g,~\beta_f\}$ take the form
\begin{align}
\calE_X+\calE_X^{\rm const} = 0
\,,
\end{align}
where $\calE_X^{\rm const}$ is the contribution from the constraint part, while the equation of motion for $\lambda$ is
\begin{align}
\calE_{\lambda}= \lambda \calD-\calC =0
\,.
\end{align}
One can easily conclude that $\lambda=0$ is a solution to the equations of motion. When $\lambda=0$ is substituted, we find $\calE_X^{\rm const}=0$ and $\calE_{\lambda}=-\calC$. As we have explained, equations $\calE_X=0$ lead to the constraint $\calC=0$; then, the equation of motion for $\lambda$ is consistently solved. Note that this analysis does not exclude the existence of other solutions, but the other solution does not work well, at least in the isotropic Universe (see Appendix~\ref{sec:hamilton_formalism}). 
Hence, the background equations of motion in MTBG are reduced to those of HRBG.

\subsection{Structure of equations of motion}
\label{sec:structure}
By the use of the freedom of the time reparametrization, $t \to t'(t)$, we impose the gauge condition $N_g=1$. The independent equations of motion are
\begin{align}
\calE_{N_g}&=0\,, \quad \calE_{N_f}=0\,, \quad \calC = 0
\,, \label{eoms:constraints} \\
\calE_{\beta_g}&=0 \,, \quad \calE_{\beta_f}=0
\,, \label{eoms:dynamical}
\end{align}
which determine the dynamics of the five variables $\{N_f, a_g, a_f, \beta_g, \beta_f \}$. The equations in \eqref{eoms:constraints} are understood as constraints since they do not contain second derivatives whereas \eqref{eoms:dynamical} are the equations of motion for the anisotropies.

To solve the equations \eqref{eoms:constraints} and \eqref{eoms:dynamical}, it is convenient to regard $\{H_g,H_f,\xi, r, \beta_g, \beta_f\}$ as independent variables. The equations \eqref{eoms:constraints} and \eqref{eoms:dynamical} are closed within $\{H_g,H_f,\xi, r, \beta_g, \beta_f\}$. However, while there are six variables, only five equations exist and an additional equation is required. The time derivative of $\xi=a_f/a_g$ is expressed as
\begin{align}
    \dot{\xi}=-\xi H_g + r \xi^2 H_f \,.
\label{eq:dotxi}
\end{align}
By taking the time derivative of $\calC=0$ and using the equations \eqref{eq:dotHg}-\eqref{eq:anisotropyf} and \eqref{eq:dotxi}, we obtain
\begin{align}
\dot{\calC}=\dot{\calC}(H_g,H_f,\xi,r,\beta,\dot{\beta}_g,\dot{\beta}_f) = 0
\,.
\end{align}
Hence, we have six equations
\begin{align}
\calE_{N_g}=0, ~ \calE_{N_f}=0, ~ \calC=0, ~ \dot{\calC}=0, ~ \calE_{\beta_g}=0, ~ \calE_{\beta_f}=0
\,,
\label{allequations}
\end{align}
which are closed within the six variables $\{ H_g,H_f,\xi, r, \beta_g, \beta_f \}$. Once the solutions to \eqref{allequations} are found, the dynamics of $\{a_g, a_f, N_f \}$ can be solved by using $\dot{a}_g = H_g a_g$ and \eqref{eq:defxir}. 

The variables $\{H_g, H_f, \xi, r\}$ are algebraically determined by $\{\beta_g,\beta_f, \dot{\beta}_g, \dot{\beta}_f \}$, although explicit solutions cannot be found due to the nonlinearity of the constraints. We only consider a branch such that $H_g>0$, namely the expanding universe. The variables $\beta_g$ and $\beta_f$ obey a couple of second-order differential equations \eqref{eoms:dynamical}. The present system requires $2\times 2$ initial conditions for integration, corresponding to one physical degree of freedom of the massless graviton and that of the tensor mode of the massive graviton, respectively. 
The equations \eqref{eoms:dynamical} give 
\begin{align}
\dot{\Sigma}_{0}+3 H_g \Sigma_{0} = 0\,,
    \label{eq:sigmaml}
\end{align}
where
\begin{align}
    \Sigma_{0} :=  \sigma_g + \alpha^{2} \xi^3 \sigma_f
    \,.
\end{align}
The solution to \eqref{eq:sigmaml} is immediately found to be $\Sigma_0 \propto a_g^{-3}$, which is the same as the decaying law of the shear in GR. Hence, $\Sigma_0$ can be interpreted as the massless mode of the shear. On the other hand, $\beta_g$ and $\beta_f$ always appear in the equations of motion in the combination $\beta=\beta_g-\beta_f$ which can be interpreted as the massive mode. (Hence, the number of physically meaningful initial conditions is $3$ rather than $4$. The redundant initial condition is the freedom associated with the global rescaling of the spatial coordinates, $x\to e^{2c} x, y \to e^{-c}y, z \to e^{-c}z$, with a constant parameter $c$.) However, the differential equation for $\beta$ cannot be expressed in a simple form. 

\subsection{Fixed points}
As explained above, the equations are nonlinear differential equations and their generic properties are not easily deduced. Therefore, by following Ref.~\cite{Gumrukcuoglu:2012aa}, we first look for solutions
under the condition
\begin{align}
    \ddot{\beta}_g=\ddot{\beta}_f=\dot{\beta}_g=\dot{\beta}_f=0\,.
    \label{eq:fixedcondition}
\end{align}
Since $\{H_g, H_f, \xi, r\}$ are determined by the algebraic equations, the above condition \eqref{eq:fixedcondition} implies
\begin{align}
\dot{H}_g=\dot{H}_f=\dot{\xi}=\dot{r}=0
\,,
\label{eq:fixedcondition2}
\end{align}
and then all the variables $\{ H_g,H_f,\xi, r, \beta_g, \beta_f \}$ remain constant. Hence, the condition \eqref{eq:fixedcondition} yields fixed-point solutions. At the fixed points, the $g$- and the $f$-spacetime themselves are isotropic because of the absence of the shear while the ratio $g^{\mu\alpha}f_{\alpha\nu}$ is anisotropic when $\beta \neq 0$. We call solutions with $\beta=0$ isotropic fixed points and those with $\beta\neq 0$ anisotropic fixed points, respectively. 

Under the fixed point conditions~\eqref{eq:fixedcondition} and \eqref{eq:fixedcondition2}, both equations for the anisotropy \eqref{eq:anisotropy} and \eqref{eq:anisotropyf} are reduced to the same equation
\begin{align}
    (e^\beta-e^{-2\beta})\left[b_1+b_2 \left(e^{\beta }+r\right)\xi+b_3 e^{\beta} r\xi^2\right]&=0\,,
    \label{eqanisotropy}
\end{align}
while the Friedmann equation for the $g$-metric \eqref{eq:Friedmanng}, that for the $f$-metric \eqref{eq:Friedmannf}, and the constraint equation become respectively
\begin{align}
    &-3 h_g^2+b_0+b_2 e^{-\beta} \left(e^{3 \beta }+2\right)  \xi ^2 
    \nn
    &\quad+b_1 \left(e^{-2 \beta }+2 e^{\beta }\right)  \xi +b_3 \xi ^3 = 0\,,
    \label{Feqg}
    \\
    &b_1+\left[ b_2 \left(e^{-2 \beta }+2 e^{\beta }\right)-3 \alpha ^2 r^{-2}h_g^2\right]\xi
    \nn
    &\quad+ b_3 \xi ^2 \left(e^{2 \beta }+2e^{-\beta}\right) + b_4 \xi ^3 = 0\,, 
    \label{Feqf}
    \\
    &b_3 \left[-3 +r\left( 2e^{-\beta }+e^{2 \beta }\right) \right]\xi^2
    \nn
    &\quad-2 b_2 \left[(2 e^{-\beta }+e^{2 \beta })-r\left( 2 e^{ \beta } +e^{-2\beta}\right)\right]\xi
    \nn
    &\quad-b_1 \left(e^{-2 \beta }+2 e^{\beta }-3 r\right)=0\,,
    \label{Ceq}
\end{align}
where we have defined a dimensionless combination $h_g:=H_g/m_g$ and have used the relations
\begin{align}
    H_f=\frac{H_g}{r\xi} \,, \quad N_f=r\xi\,,
\end{align}
following from \eqref{eq:defxir} and \eqref{eq:fixedcondition2}.

We first consider the isotropic case $\beta=0$. The constraint equation \eqref{Ceq} is reduced to
\begin{align}
    (r-1)(b_1 +2 b_2 \xi +b_3 \xi^2) = 0\,.
    \label{eq:isotropic_fixed_point}
\end{align}
This equation has two branches. The first branch $r=1$ is called the {\it normal branch} and it leads to the relation $H_f=\xi H_g$. In this branch, eliminating $h_g$ from the Friedmann equations \eqref{Feqg} and \eqref{Feqf}, and using $\beta=0$, we obtain
\begin{align}
    &-\alpha ^2 b_3 \xi ^4+\left(4 b_4-3 \alpha ^2 b_2\right)\xi ^3 +3  \left(b_3-\alpha ^2 b_1\right)\xi ^2 
    \nn
    &+  \left(3 b_2-\alpha ^2
   b_0\right)\xi+b_1 = 0\,,
   \label{eq:xi_normal_branch}
\end{align}
which is an algebraic equation for $\xi$ and can be solved for $\xi$. Substituting the root $\xi$ into the Friedmann equation \eqref{Feqg}, the Hubble parameter $h_g$ is fixed in terms of the coupling constants of the theory.
On the other hand, the second branch $b_1 +2 b_2 \xi +b_3 \xi^2=0$ is called the {\it self-accelerating branch}. By using a root of $b_1 +2 b_2 \xi +b_3 \xi^2=0$, the Hubble parameter $h_g$ and the ratio $r$ are determined by \eqref{Feqg} and \eqref{Feqf}. In particular, $r$ is given by
\begin{align}
   r = \alpha\sqrt{\frac{\xi \left(b_3 \xi^3+3 b_2 \xi^2+3 b_1 \xi+b_0\right)}{b_4 \xi^3 + 3 b_3 \xi ^2+3 b_2 \xi+b_1}}\,.
   \label{eq:r_selfaccelerating}
\end{align}
In the case of HRBG, the self-accelerating branch would suffer from a nonlinear instability as with the dRGT theory~\cite{DeFelice:2012mx}. The normal branch is stable when the Hubble parameter is sufficiently small while the scalar mode of the massive graviton becomes a ghost, known as the Higuchi ghost, when the Hubble parameter exceeds a critical value~\cite{Higuchi:1986py,Higuchi:1989gz} (see also~\cite{Grisa:2009yy,Comelli:2012db,DeFelice:2014nja,Aoki:2015xqa}). On the other hand, MTBG can avoid both instabilities thanks to the absence of the dynamical scalar mode~\cite{DeFelice:2020ecp}.

Next, we consider the anisotropic fixed points, $\beta\neq 0$. Eliminating $b_3$ from \eqref{eqanisotropy} and \eqref{Ceq}, we obtain
\begin{align}
    (1-e^{\beta})\left(b_2 e^{\beta } \xi +b_1\right)\left(r-e^{\beta}\right) \left(r-e^{-2 \beta} \right) =0\,.
    \label{eq:EconEani}
\end{align}
where $-3+r(e^{-2\beta}+2e^{\beta}) \neq 0$ is assumed. Note that the isotropic limit $\beta \to 0$ leads to $-3+r(e^{-2\beta}+2e^{\beta}) \to -3(1-r)$ so the anisotropic extension of the normal branch does not have to satisfy \eqref{eq:EconEani}. There are in principle four ways to satisfy \eqref{eq:EconEani}, defining up to four different branches. The branch $e^{\beta}=1$ corresponds to the isotropic self-accelerating branch while the other three branches,
\begin{align}
    e^{\beta} = \left\{ -\frac{b_1}{b_2 \xi}\,,~r\,,~r^{-1/2}  \right\}\,,
   \label{aniso_beta}
\end{align}
may lead to anisotropic fixed points. As in the dRGT theory~\cite{Gumrukcuoglu:2012aa}, either $e^\beta=-b_1/(b_2\xi)$ or $e^{\beta}=r$ does not give interesting solutions, and non-trivial anisotropic fixed points can be found in the third branch $e^{\beta}=r^{-1/2}$. In the following, we discuss them in order.

\underline{Branch 1.} ~ Substituting the solution $e^\beta=-b_1/(b_2\xi)$ into \eqref{eqanisotropy}, we obtain
\begin{align}
    \left(b_2^3 \xi^3+b_1^3\right)\left(b_2^2-b_1 b_3\right) = 0\,.
\end{align}
The first solution $\xi=-b_1/b_2$ gives $\beta=0$ and thus this solution is not anisotropic. The second solution $b_2^2 - b_1 b_3 = 0$ requires a parameter tuning. In this case, the equations of motion yield
\begin{align}
    \xi r = \sqrt{\frac{\alpha^2 b_2(b_2^3-b_0 b_3^2)}{b_3^2 (b_3^2-b_2 b_4)}}\,,
    \\
    h_g = \sqrt{\frac{\alpha^2(-b_2^3+b_0 b_3^2)}{3 b_3^2 (1+\alpha^{2})}}\,,
    \\
    h_f = \sqrt{\frac{b_2 b_4 -b_3^2}{b_2(3\alpha^2+1)}}\,,
    \\
    \frac{e^{\beta}}{r}=\sqrt{\frac{b_2 \left(b_3^2-b_2 b_4\right)}{\alpha^2 (b_2^3-b_0 b_3^2)}}\,,
\end{align}
where the variables $r, \xi$ and $\beta$ are not completely determined, that is, the fixed point is not isolated. Therefore, we shall not discuss this branch furthermore.

\underline{Branch 2.} ~ We then consider the solution $r=e^{\beta}$. Substituting this into the equation for anisotropy \eqref{eqanisotropy}, we obtain
\begin{align}
    b_3 e^{2\beta} \xi^2 +2 b_2e^{\beta } \xi+b_1 = 0\,,
\end{align}
which is solved by
\begin{align}
    \xi = \frac{e^{-\beta } \left( -b_2 \pm \sqrt{b_2^2-b_1 b_3}\right)}{b_3}
    \,.\label{eq:xibranch3}
\end{align}
Then the Friedmann equations \eqref{Feqg} and \eqref{Feqf} give
\begin{align}
    h_g^2&=\frac{2 b_2^3-3 b_1 b_2 b_3+b_1^2 b_4 \pm 2(b_2^2-b_1b_3)^{3/2}}{3 b_3^2}\,,
    \\
    0&= -2 b_2^2b_4+b_2 b_3^2+b_1 b_3 b_4+\alpha^2\left(2 b_2^3-3 b_1 b_2 b_3+b_0 b_3^2\right) 
    \nn
    &\pm 2 \sqrt{b_2^2 - b_1 b_3} \left[ \left(b_3^2-b_2 b_4\right)+\alpha^{2}\left(b_2^2-b_1 b_3\right)\right] \,.
\end{align}
The first equation determines the Hubble parameter in terms of the coupling constants while the second equation imposes a constraint on the coupling constants rather than determining the value of $e^{\beta}$. Hence, this branch is not of our interest.

\underline{Branch 3.} ~ Finally, we discuss the third branch $r=e^{-2\beta}$. With this solution, the anisotropy equation \eqref{eqanisotropy} and a combination of \eqref{Feqg} and \eqref{Feqf} gives algebraic equations for $\xi$ and $e^{\beta}$:
\begin{align}
    &b_3 e^{-\beta}\xi^2+b_2(e^{-2\beta}+e^{\beta})\xi+b_1=0\,.
    \label{fixedanisotropiceq}
    \\
    &b_3 \alpha^2 \xi^4+\left[b_2\alpha^2(2e^{-\beta}+e^{2\beta})-b_4 e^{-4 \beta}\right]\xi^3
    \nn
    &+ \left[ b_1 \alpha^2 (e^{-2\beta}+2e^{\beta}) -b_3 (2 e^{-5 \beta}+e^{-2\beta}) \right]\xi^2
    \nn
    &+ \left[ b_0 \alpha^2 -b_2 (e^{-6 \beta}+2 e^{-3\beta}) \right] \xi -b_1 e^{-4\beta} = 0 \,,
    \label{eq:xipoly}
\end{align}
We can further combine \eqref{fixedanisotropiceq} and \eqref{eq:xipoly} to find an expression linear in $\xi$,
\begin{align}
\xi =- \frac{ b_1 \left[ b_3^2-b_2b_4 + \alpha^2  (b_2^2-b_1 b_3) e^{3\beta} \right] (e^{2 \beta }+e^{5\beta }) }{\calQ_0 + \calQ_1 e^{3\beta} + \calQ_2 e^{6\beta} +\calQ_3 e^{9\beta}
      }
\,, \label{eq:solxi}
\end{align}
and a quartic-order algebraic equation for $e^{3\beta}$,
\begin{align}
    \calC_0 + \calC_1 e^{3\beta} + \calC_2 e^{6\beta} + \calC_3 e^{9\beta} + \calC_4 e^{12\beta} = 0 \,,
    \label{eq:betaequation}
\end{align}
where the coefficients are given by
\begin{align}
    \calQ_0 &= b_2 (b_3^2-b_2 b_4)\,, \\
    \calQ_1 &= b_2 b_3^2-2b_2^2 b_4 + b_1 b_3 b_4 + \alpha^2b_2(b_2^2-b_1 b_3) \,, \\
    \calQ_2 &= b_2(b_3^2-b_2 b_4) +\alpha^2 (2b_2^3-3b_1 b_2 b_3 + b_0 b_3^2)\,, \\
    \calQ_3 &= \alpha^2 b_2 (b_2^2-b_1 b_3) 
    \,,
\end{align}
and
\begin{align}
    \calC_0 &= \left(b_2^2-b_1 b_3\right) \left(b_3^2-b_2 b_4\right)\,,
    \\
    \calC_1 &=  -2 b_4 b_2^3+b_3^2 b_2^2+4 b_1 b_2 b_3 b_4-2 b_1 b_3^3-b_1^2 b_4^2
    \nn
    &+\alpha^2(b_2^4-2 b_1 b_2^2 b_3+b_0 b_2^2 b_4 -b_0 b_2 b_3^2 -b_1^2 b_2 b_4 +2 b_1^2 b_3^2)
    \,,
    \\
    \calC_2 &=  \left(b_2^2-b_1 b_3\right) [ b_3^2-b_2 b_4+\alpha^{2}\left(2 b_2^2-4 b_1 b_3+2 b_0 b_4\right)
    \nn
    &+\alpha^4 (b_1^2-b_0 b_2) ]\,,
    \\
    \calC_3 &= \alpha^{2}( b_2^4-2 b_1 b_2^2 b_3+b_0 b_2^2 b_4-b_0 b_2 b_3^2-b_1^2 b_2 b_4+2 b_1^2 b_3^2)
    \nn
    &+\alpha^{4}(-2 b_1^3 b_3+b_1^2 b_2^2 +4 b_0 b_1 b_2 b_3 -2 b_0 b_2^3-b_0^2 b_3^2) \,,
    \\
    \calC_4&= \alpha^4 \left(b_1^2-b_0 b_2\right) \left(b_2^2-b_1 b_3\right)\,.
\end{align}
Since \eqref{eq:betaequation} is quartic order in $e^{3\beta}$, there are four independent roots of the algebraic equation, in general. Once a root is chosen, $\xi$ and $h_g^2$ are uniquely determined by \eqref{eq:solxi} and \eqref{Feqg}. Hence, unlike the other branches, all the variables $\{r, \xi, \beta, h_g\}$ are fixed without any fine-tuning of the coupling constants. We thus focus on this branch in the following.

\section{Stability of fixed points}
\label{sec:stability}
In this section, we study the stability of the fixed points obtained in the previous section. 
As we have explained, the system involves one massless degree of freedom and one massive degree of freedom. In particular, the massless mode $\Sigma_0$ decays as $a_g^{-3}$ and can be ignored as the universe expands. Since our interest is in the dynamics of the massive mode, we shall assume
\begin{align}
 \Sigma_0 = \sigma_g + \alpha^{2} \xi^3 \sigma_f = 0
 \,.
\end{align}
in which the dimension of the phase space is reduced to two. In principle, the equations of motion can be reduced to a single second-order differential equation for $\beta=\beta_g-\beta_f$ (or a couple of first-order differential equations) when the constraints are solved. In practice, however, the constraints are nonlinear and cannot be solved explicitly. Hence, we classify the fixed points based on the stability against small perturbations by which the equations are linearized. The global stability is then examined by using two-dimensional phase portraits.

\subsection{Local stability}

The equations \eqref{eq:anisotropy} and \eqref{eq:anisotropyf} yield
\begin{align}
0=\calE_{\beta}&:=
    \ddot{\beta}+\frac{( r+3 \alpha^2 \xi^2)H_g+2 r^2 \xi  H_f - \dot{r}}{ r+\alpha^2\xi^2}\dot{\beta}
    \nn
    &+\frac{m^2}{3(1+\alpha^2)\xi} (e^{\beta}-e^{-2\beta})(r +\alpha^2\xi^2)
    \nn
    &\quad\times[b_1 +b_2(e^{\beta}+r)\xi+b_3 e^{\beta}r\xi^2] \,.
    \label{eq:sigmams}
\end{align}
where $\Sigma_0=0$ is used. We consider perturbations around the fixed points as
\begin{align}
    H_g &= m_g (h_{g0}+\epsilon h_{g1}(t))\,,
    \label{eq:hg_pert}
    \\
    H_f &= m_g \left( \frac{h_{g0}}{r_0 \xi_0} +\epsilon h_{f1}(t)\right)\,,
    \label{eq:hf_pert}
    \\
    \xi &= \xi_0 + \epsilon \xi_1(t)\,,
    \label{eq:xi_pert}
    \\
    r &= r_0+\epsilon r_1(t)\,,
    \label{eq:r_pert}
    \\
    \beta &= \beta_0+\epsilon \beta_1(t)\,.
    \label{eq:beta_pert}
\end{align}
Here, the quantities with the subscript $0$ are the fixed-point solutions which are determined in terms of the coupling constants while $\{ h_{g1},h_{f1},\xi_1,r_1,\beta_1\}$ represent the perturbations and we have introduced a small parameter $\epsilon$ to keep track of orders of parturbations. The linearized equation for $\beta$ is given by
\begin{align}
\calE^{(1)}_{\beta}=\ddot{\beta}_1 + 3H_{g0} \dot{\beta}_1
+\calE^{(1)}_{\beta\beta}\beta_1 + \calE^{(1)}_{\beta r} r_1 + \calE^{(1)}_{\beta \xi} \xi_1
=0
\,,
\end{align}
where $H_{g0}=m_g h_{g0}$ and the coefficients $\calE^{(1)}_{\beta\beta}, \calE^{(1)}_{\beta r}, \calE^{(1)}_{\beta \xi} $ are computed for each fixed-point solution. As we have explained, $\{H_g, H_f, \xi, r \}$ are fixed by the constraints. Thanks to the linearization, the constraints can be explicitly solved for $h_{g1},h_{f1},\xi_1,r_1$ although the exact expressions are lengthy. We then obtain a second-order differential equation for $\beta_1$.

In the case of the isotropic fixed points, $\beta_0=0$, the coefficients $\calE^{(1)}_{\beta r}$ and $\calE^{(1)}_{\beta \xi} $ vanish and then we do not need to solve the constraints explicitly. The equation for $\beta_1$ is given by
\begin{align}
\ddot{\beta}_1 + 3H_{g0} \dot{\beta}_1 + M^2_I \beta_1 =0
\,.
\end{align}
with
\begin{align}
    M_{I}^2 &= \frac{m^2 \left( r_{0}+\alpha^2\xi_{0}^2 \right) \left[b_1+b_2 \xi _{0} \left(r_{0}+1\right)+b_3 r_{0} \xi_{0}^2 \right] }{(1+\alpha^2) \xi _{0}}\,.
    \label{eq:mass_iso}
\end{align}
The values of $r_0$ and $\xi_0$ are fixed by choosing the branch: $\xi_0$ is a root of \eqref{eq:xi_normal_branch} and $r_0=1$ in the normal branch while $\xi_0$ is a root of $b_1 +2 b_2 \xi +b_3 \xi^2=0$ and $r_0$ is given by \eqref{eq:r_selfaccelerating} in the case of the self-accelerating branch, respectively.

At the anisotropic fixed points, on the other hand, the coefficients $\calE^{(1)}_{\beta r}$ and $\calE^{(1)}_{\beta \xi} $ do not vanish and the constraints need to be solved. We recall that the anisotropic fixed points satisfy
\begin{align}
r_0 = e^{-2\beta_0}\,, \quad
b_3 e^{-\beta_0}\xi_0^2+b_2(e^{-2\beta_0}+e^{\beta_0})\xi_0+b_1=0
\,, \label{anisoFP}
\end{align}
which can be used to simplify the expressions. Using \eqref{anisoFP} to eliminate $r_0$ and $b_1$, we finally obtain
\begin{align}
\ddot{\beta}_1 + 3H_{g0} \dot{\beta}_1 + M^2_A \beta_1 =0
\,.
\end{align}
where the mass squared is given by
\begin{align}
    M_{A}^2 = \frac{m^2 d_1 d_2 d_3 e^{-5\beta_0} [- d_1 d_2 + 6 \alpha^2 e^{6\beta_0} h_{g0}^2 ]}{(1+\alpha^2) [ d_1 d_2^2 +2 \alpha^2 e^{6\beta_0}  h_{g0}^2  (3d_2+2 d_3 e^{\beta_0}) ]},
\end{align}
with
\begin{align}
    d_1 &:= \left(e^{3 \beta_0 }-1\right) \left(1 +\alpha^2e^{2 \beta_0 } \xi_0 ^2\right)\,,
    \\
    d_2 &:= e^{\beta_0 } b_3 \xi_0 +b_2\,,
    \\
    d_3 &:= b_2 e^{2\beta_0}+b_3\xi_0\,.
\end{align}

\begin{table*}[t]
  \centering
  \begin{tabular}{|c||c|c|c|}
    \hline
        &\begin{tabular}{c}
        stable spiral \\
        (damped-oscillation)
    \end{tabular} 
    &\begin{tabular}{c}
        stable node \\
        (over-damping)
    \end{tabular} 
    &\begin{tabular}{c}
        saddle point \\
        (unstable)
    \end{tabular} 
    \\\hline
    $M^2$  & $+$  &  $+$ & $-$  \\
    \hline 
    $9 H_{g0}^2-4M^2$  & $-$  & $+$ &  $+$ \\
    \hline
    phase portraits
    &\begin{tabular}{c}
        isotropic: Fig. \ref{fig:isotrpic_stable_spiral} \\
        anisotropic: Fig. \ref{fig:anisotrpic_stable_spiral}
    \end{tabular}
    &\begin{tabular}{c}
        isotropic: Fig. \ref{fig:isotrpic_stable_node}\\
        anisotropic: Fig. \ref{fig:anisotrpic_stable_node} 
    \end{tabular} 
    &\begin{tabular}{c}
        isotropic: Fig. \ref{fig:isotrpic_saddle} \\
        anisotropic: Fig. \ref{fig:anisotrpic_saddle}
    \end{tabular} 
    \\\hline
  \end{tabular}
  \caption{Classification of the fixed points.}
  \label{table:calssification}
\end{table*}

Therefore, in either case, the linearized equation for $\beta$ takes the form
\begin{align}
\ddot{\beta}_1 + 3H_{g0} \dot{\beta}_1 + M^2 \beta_1 =0
\,,
\label{eq:linearlizedequation}
\end{align}
where $M^2$ is either $M_I^2$ (isotropic fixed points) or $M_A^2$ (anisotropic fixed points). 
This equation is consistent with the linear equation of the tensor modes of the massive graviton as long as the gradient term is ignored at least around the isotropic fixed point. Thus, the masses $M$ are considered as the graviton mass since \eqref{eq:linearlizedequation} is identical to the superhorizon limit of the linear equation of the tensor modes at least around the isotropic fixed point. 

We then split the second-order differential equation \eqref{eq:linearlizedequation} into a couple of first-order differential equations:
\begin{align}
\dot{\bm{v}} = K \bm{v}\,, \quad 
\bm{v}=
\begin{pmatrix}
\Sigma_{{\rm m}1} \\
\beta_1
\end{pmatrix}
\end{align}
with
\begin{align}
    K=
        \begin{pmatrix}
            -3H_{g0} & -M^2\\
            1 & 0
        \end{pmatrix}
     \,.
\end{align}
The property of the fixed points are classified by the eigenvalues of the matrix $K$
\begin{align}
    \lambda_{\pm}:=\frac{1}{2}\left(-3H_{g0} \pm \sqrt{9H_{g0}^2-4M^2}\right)
    \,,\label{eq:eigenvalue}
\end{align}
which we summarize in Table~\ref{table:calssification}.\footnote{Strictly speaking, there are other cases such as non-isolated fixed points at the boundary of the classifications. Since the fine-tuning of the coupling constants is required, we shall not discuss these cases in this paper. } Recall that we are interested in the expanding universe $H_{g0}>0$. In the case of $M^2<0$, both eigenvalues are real and satisfy $\lambda_{+}>0$ and $\lambda_{-}<0$. Therefore, a fixed point with $M^2<0$ is a saddle point. When $M^2>0$, the fixed point is locally stable because the real parts of both eigenvalues are always negative. Depending on the sign of $9H_{g0}^2-4M^2$, the stable fixed points are divided into stable spirals $(9H_{g0}^2-4M^2 < 0)$ and stable nodes $(9H_{g0}^2-4M^2 > 0)$.
The anisotropy $\beta$ is overdamping due to a large Hubble friction around the stable nodes; the eigenvalues are complex around the stable spirals and the anisotropy exhibits damped oscillation. 
All the cases can be realized in both isotropic fixed points and anisotropic fixed points when the coupling constants are appropriately chosen.

\subsection{Global stability}

\begin{figure*}
    \subfloat[$b_0=9.32,b_1=-0.0162,b_2=-0.0479,b_3=0.0122,b_4=0.00549,\alpha=1$.]{
    \includegraphics[width=0.3\linewidth]{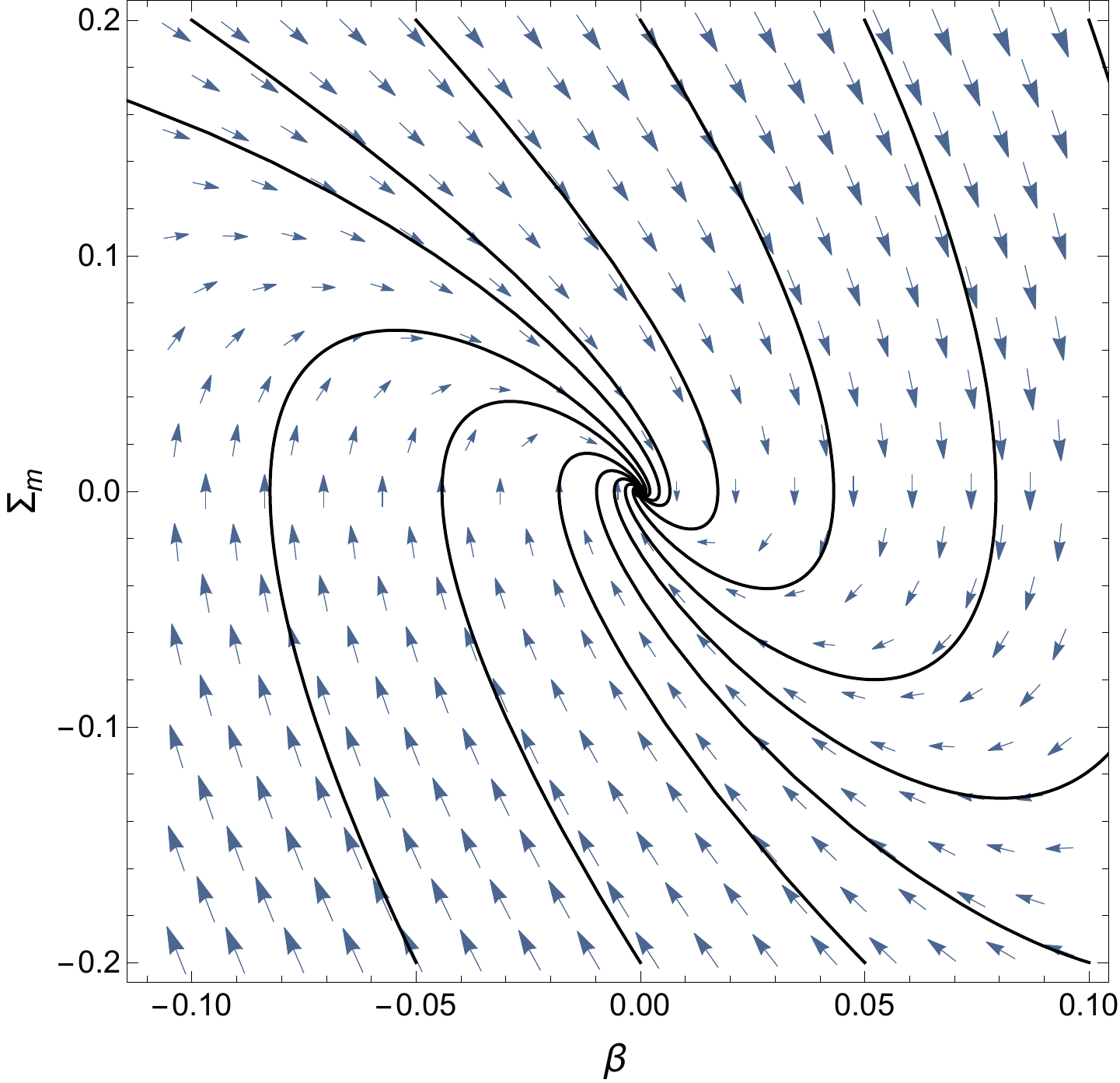}
    \label{fig:isotrpic_stable_spiral}
    }
    \subfloat[$b_0=6.61,b_1=-0.0542,b_2=-0.00258,b_3=0.00320,b_4=0.00357,\alpha=1$.]{
    \includegraphics[width=0.3\linewidth]{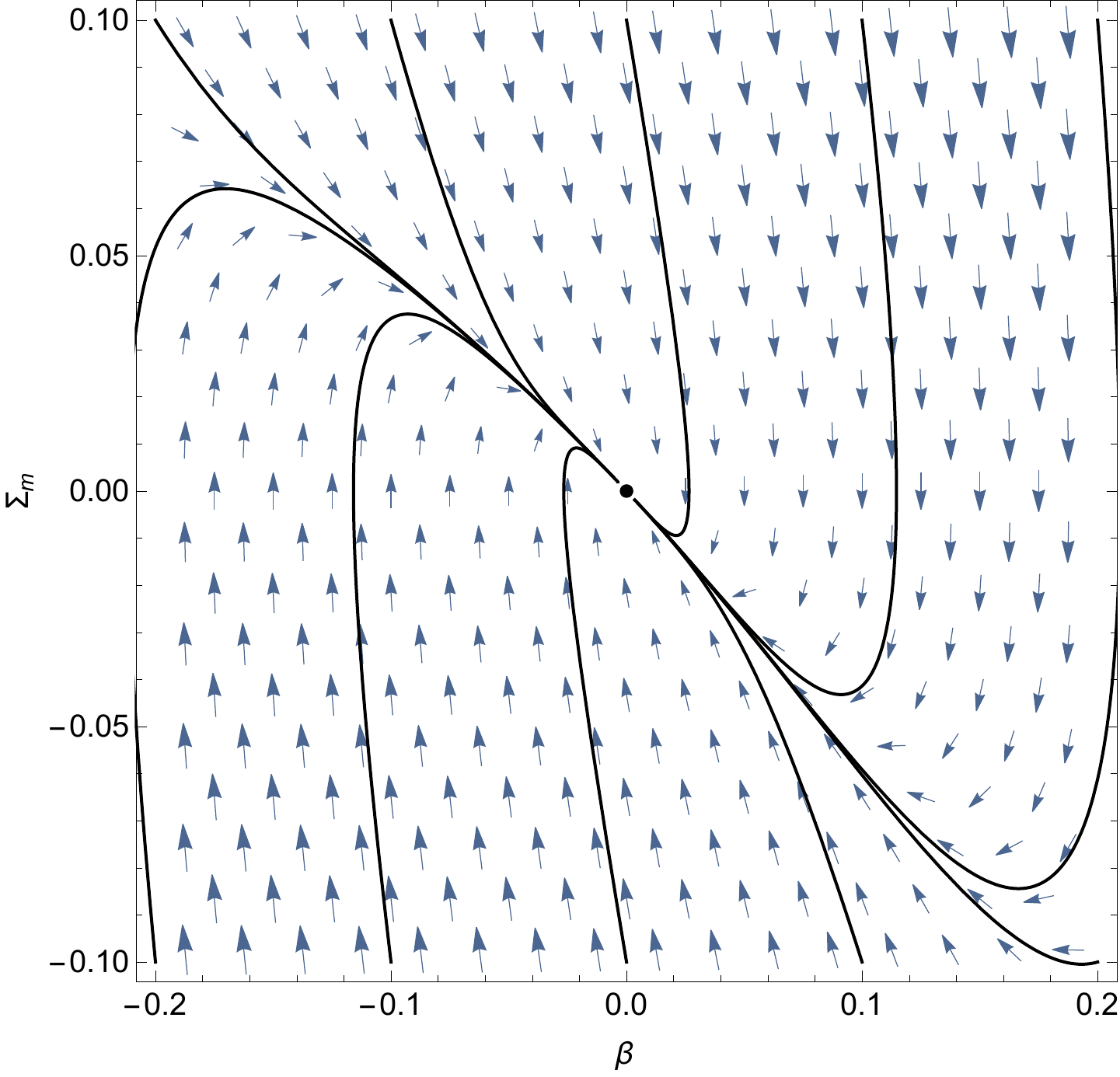}
    \label{fig:isotrpic_stable_node}
    }
    \subfloat[$b_0=50,b_1=1,b_2=8.15,b_3=-12.0,b_4=26.6,\alpha=1$.]{
    \includegraphics[width=0.3\linewidth]{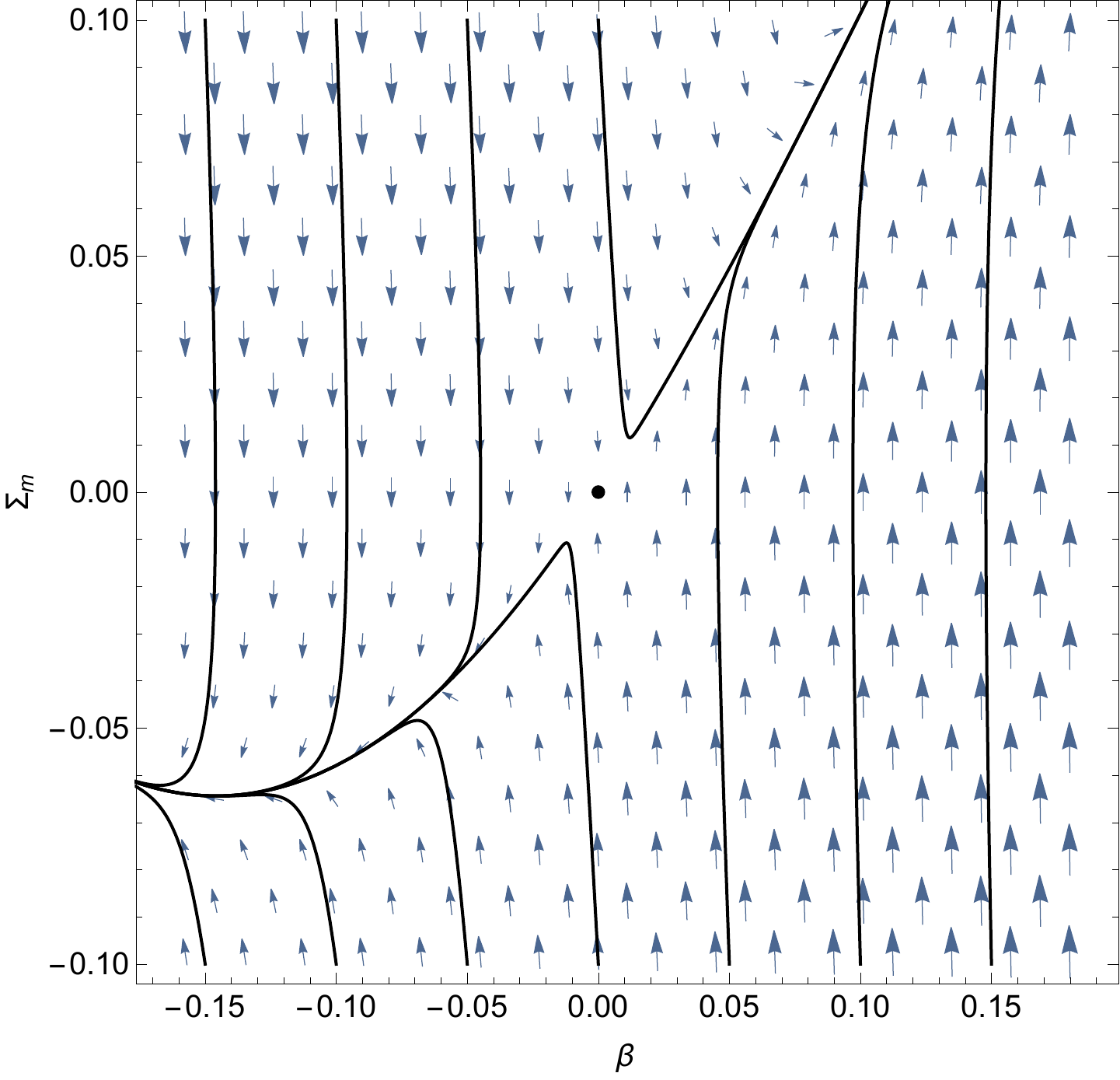}
    \label{fig:isotrpic_saddle}
    }
    \\
    \subfloat[$b_0=-6.8, b_1=4, b_2=-1.9, b_3=0.95, b_4=-1, \alpha=1$.]{
    \includegraphics[width=0.3\linewidth]{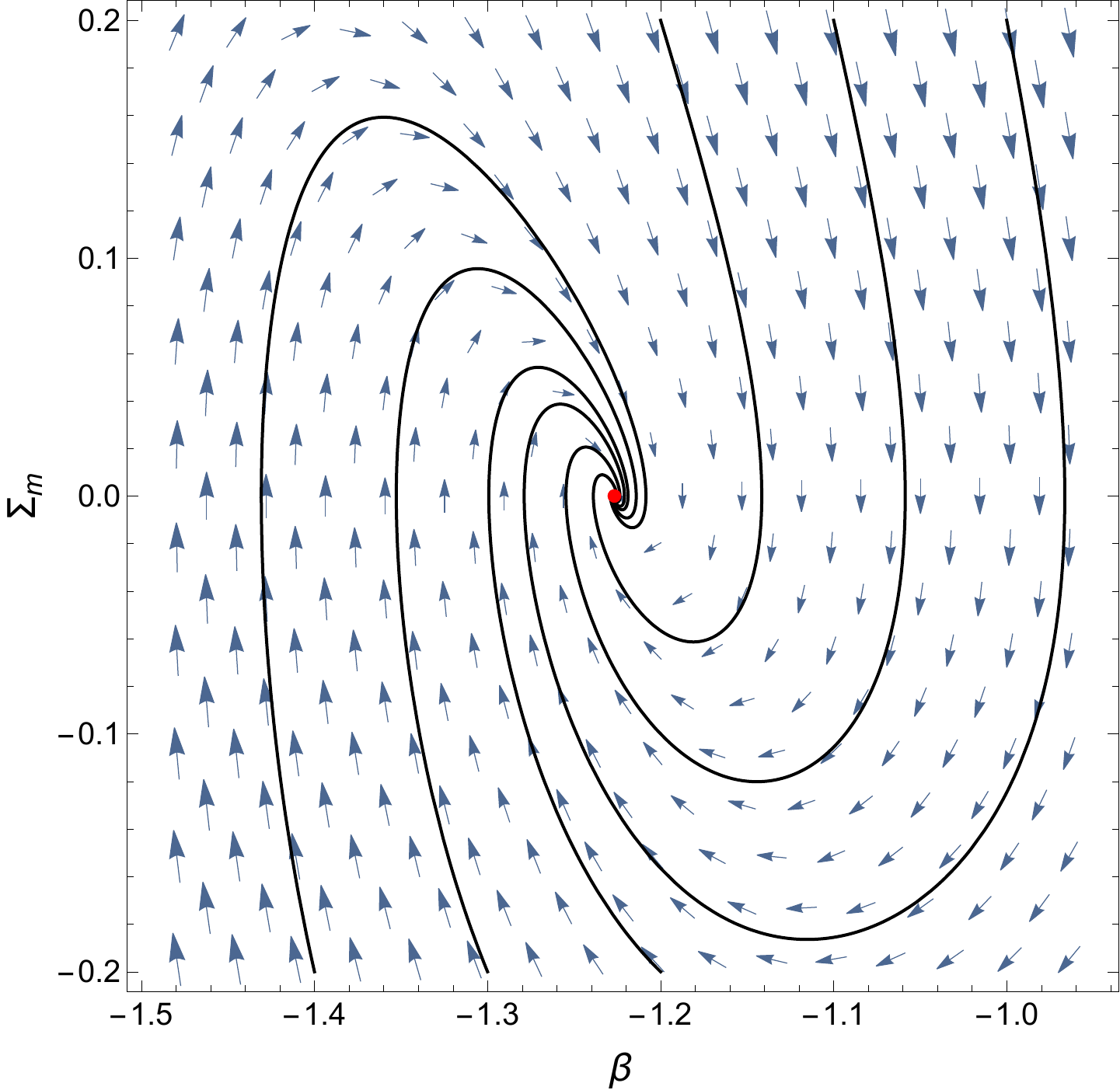}
    \label{fig:anisotrpic_stable_spiral}
    }
    \subfloat[$b_0=50,b_1=1,b_2=8.15,b_3=-12.0,b_4=26.6,\alpha=1$.]{
    \includegraphics[width=0.3\linewidth]{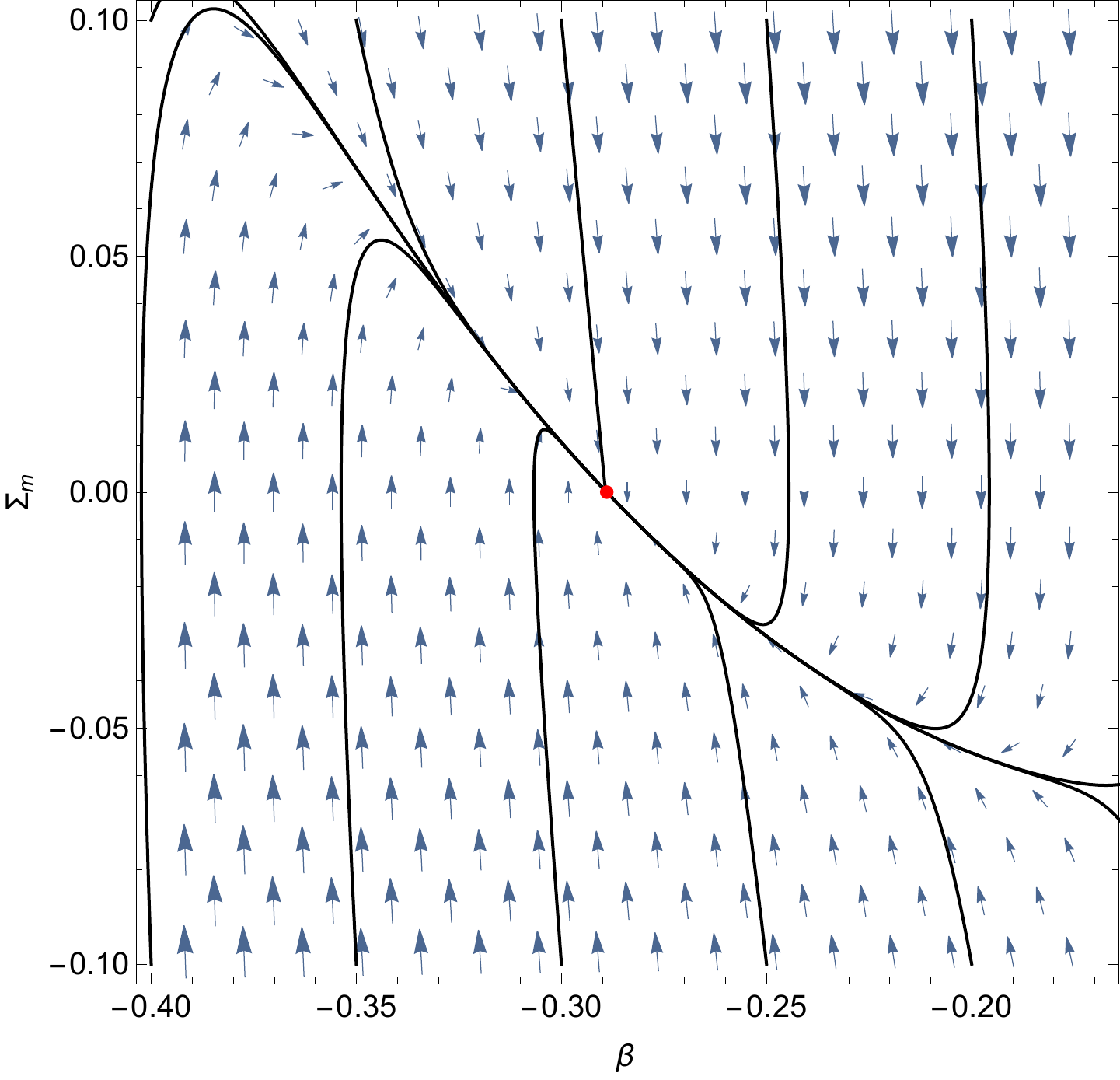}
    \label{fig:anisotrpic_stable_node}
    }
    \subfloat[$b_0=9.32,b_1=-0.0162,b_2=-0.0479,b_3=0.0122,b_4=0.00549,\alpha=1$.]{
    \includegraphics[width=0.3\linewidth]{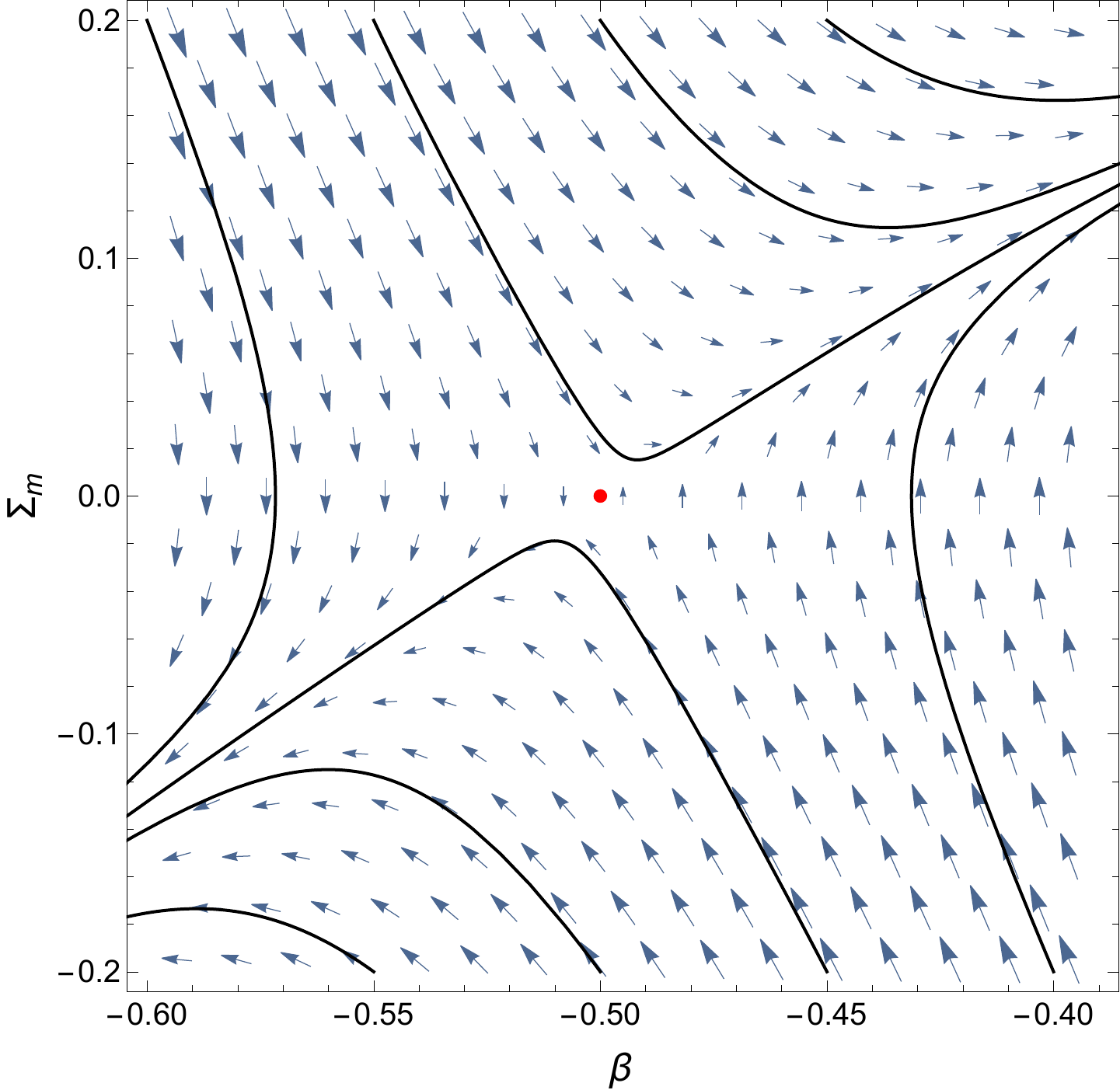}
    \label{fig:anisotrpic_saddle}
    }
\caption{Examples of phase portraits around fixed points: stable spirals (left), stable nodes (middle), and saddle points (right). The black points in the top figures represent the isotropic fixed points (self-accelerating branch) and the red points in the bottom figures are the anisotropic fixed points. The black curves are the trajectories of numerical solutions. The parameters are chosen as specified in each figure.}
\label{fig:allfixedpoints}
\end{figure*}

\begin{figure}
    \subfloat[Anisotropic stable node and isotropic saddle point:
    $b_0=50,b_1=1,b_2=8.15,b_3=-12.0,b_4=26.6,\alpha=1$.]{
    \includegraphics[width=0.9\linewidth]{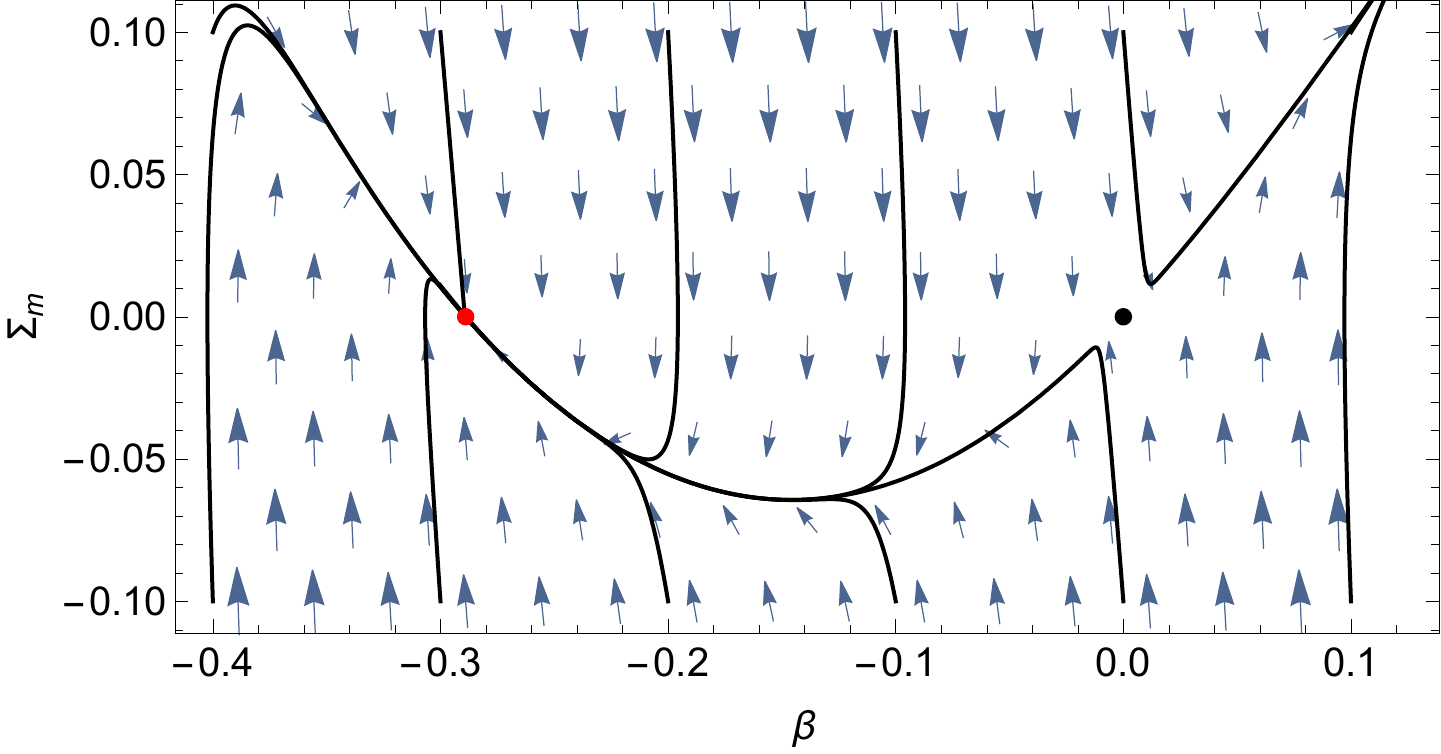}
    \label{fig:node_saddle}
    }
    \\
    \subfloat[Anisotropic saddle point and isotropic stable node: 
    $b_0=6.61,b_1=-0.0542,b_2=-0.00258,b_3=0.00320,b_4=0.00357,\alpha=1$.]{
    \includegraphics[width=0.9\linewidth]{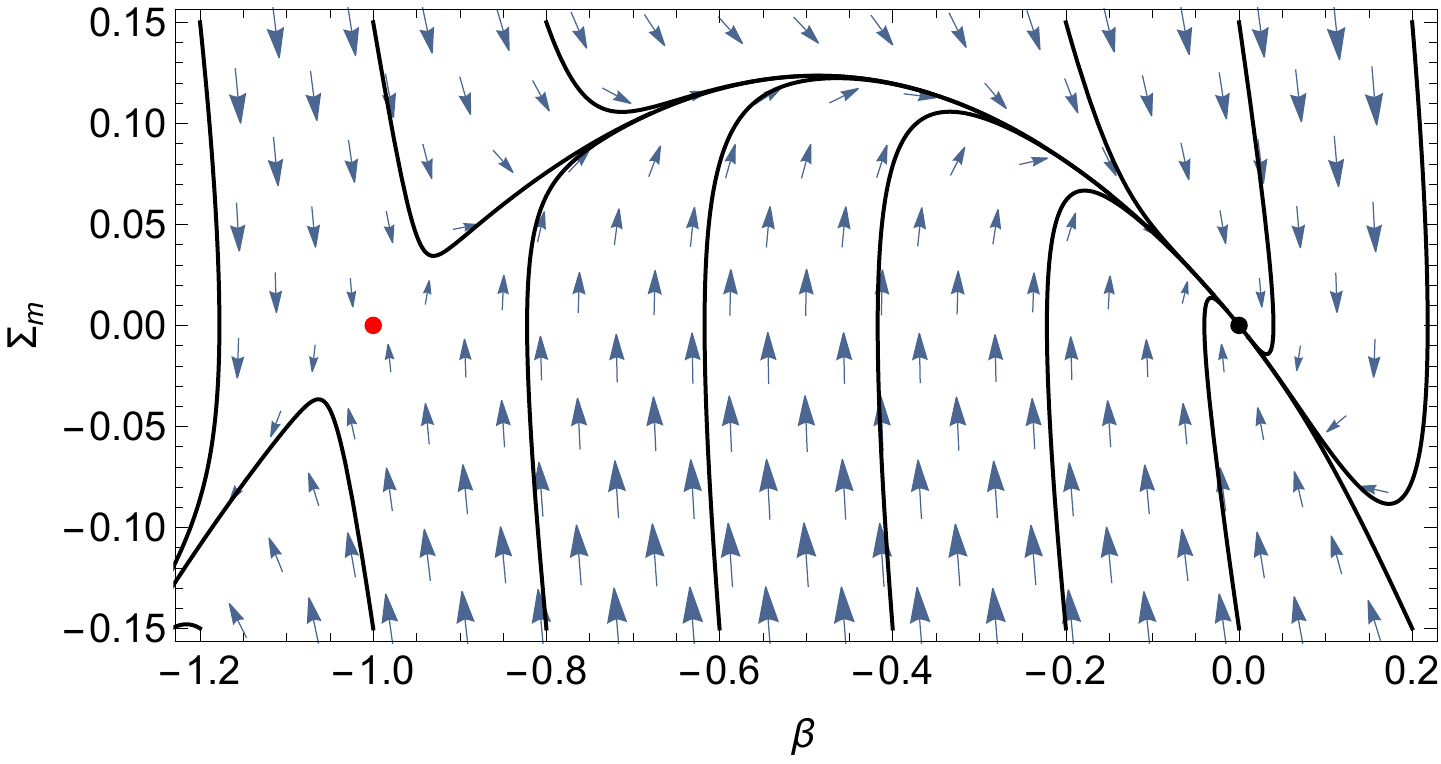}
    \label{fig:saddle_node}
    }
    \\
    \subfloat[Anisotropic saddle point and isotropic stable spiral: $b_0=9.32,b_1=-0.0162,b_2=-0.0479,b_3=0.0122,b_4=0.00549,\alpha=1$.]{
    \includegraphics[width=0.9\linewidth]{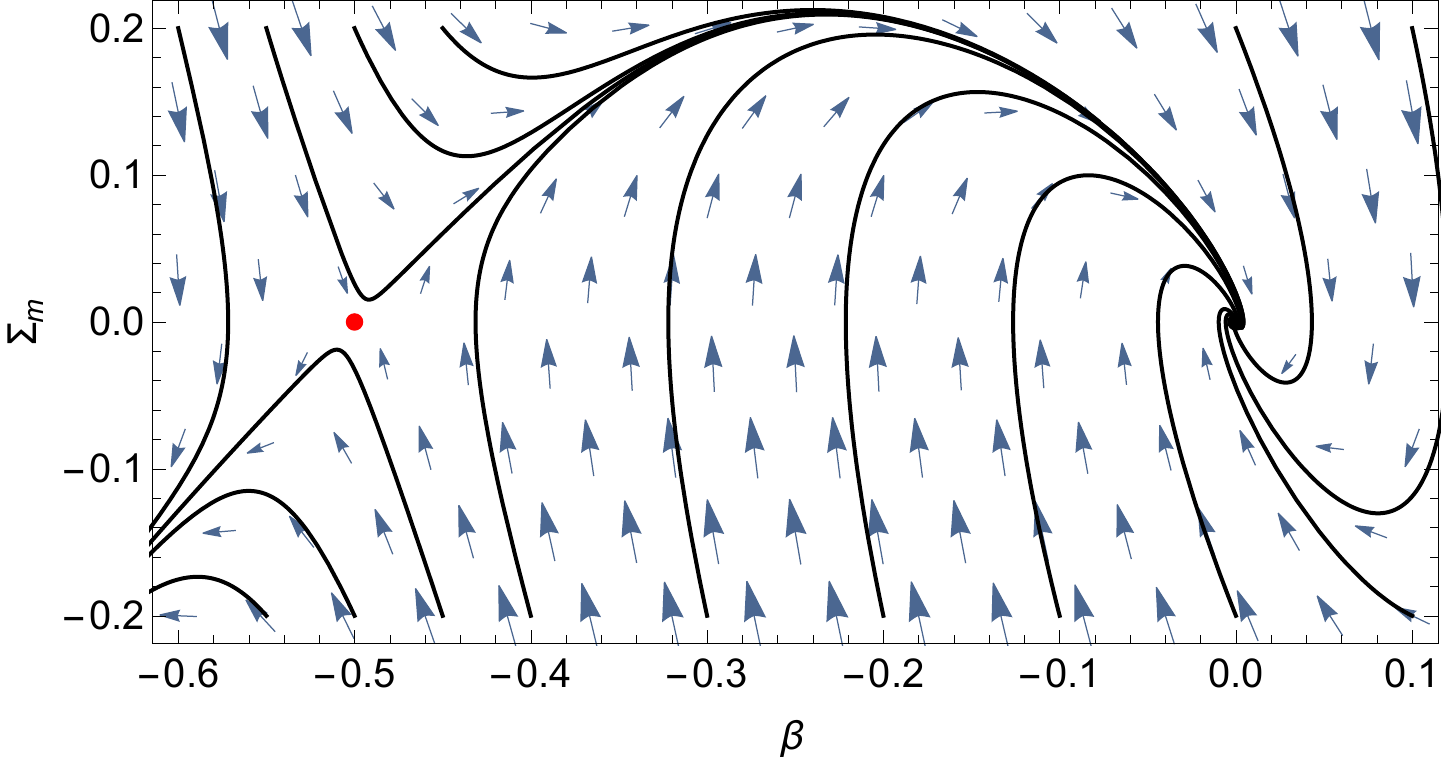}
    \label{fig:saddle_spiral}
    }
    \caption{Global structure of phase portraits. Top: the isotropic fixed point is unstable and the universe evolves towards the anisotropic fixed point. Middle: the universe generically approaches the isotropic fixed point without oscillation. Bottom: the universe moves towards the isotropic fixed point with oscillation. }
    \label{fig:global}
\end{figure}

The current universe has to be around a stable spiral to explain the dark matter by the coherent oscillation of the massive graviton. On the other hand, the initial condition is not necessarily in the vicinity of the stable spiral. Let us then discuss the global structure of the system by using phase portraits.

The set of independent equations is given in \eqref{allequations}. At each point in the phase space $(\beta, \Sigma_{\rm m})$, where $\Sigma_{\rm m}=\dot{\beta}$, their time derivatives are computed by solving \eqref{allequations} combined with the condition $\Sigma_0 =0$. However, due to the nonlinearity of the equations, there are multiple branches and we have to choose the correct branch. We first choose a fixed point and then consider the vicinity of the fixed point. The branch of the solutions in the vicinity is then chosen so that the solution is continuously connected to the fixed point, which is numerically achieved by employing the Newton-Raphson method. Iterating this procedure, we can obtain a phase portrait around each of the fixed points.

Fig.~\ref{fig:allfixedpoints} shows the phase portraits around the isotropic fixed points and the anisotropic fixed points. Although only the phase portraits of the self-accelerating branch are shown for the isotropic fixed points, similar figures can be obtained for the normal branch as well. We also integrate the equations \eqref{allequations} numerically. The trajectories of the numerical solutions are shown as the black curves in Fig.~\ref{fig:allfixedpoints}. The solutions indeed behave as classified in the perturbative analysis even at a finite distance away from the fixed point.

For a given value of the coupling constants, the equations may have several fixed points which can or cannot be connected through a dynamical evolution. We find that the anisotropic fixed point can be continuously connected to the self-accelerating branch of the isotropic fixed point. Fig.~\ref{fig:global} shows three phase portraits which exhibit flows from saddle points to stable fixed points. In Fig.~\ref{fig:node_saddle}, the isotropic universe is unstable. Even if the initial condition is isotropic, the universe typically moves towards the anisotropic fixed point when $\beta <0$. Hence, those parameters realize a spontaneous growth of the anisotropy from a tiny anisotropy. On the other hand, Figs.~\ref{fig:saddle_node} and \ref{fig:saddle_spiral} are the cases with stable isotropic universes. Although the solutions go away from the isotropic stable point if the initial value of $\beta$ is largely negative, the solutions generically approach the isotropic universe under a wide range of initial conditions. In particular, the anisotropy oscillates with a decreasing amplitude around the isotropic universe in Fig.~\ref{fig:saddle_spiral} and behaves as a dark matter component of the universe.
Therefore, when the coupling constants are appropriately chosen, the spin-2 dark matter scenario is stably realized under generic initial conditions.

\section{Dark matter production}
\label{sec:DM_production}

In the previous section, we have found that the isotropic universe can be unstable and one of the endpoints of the instability is the anisotropic fixed point. This solution may be used for a novel production mechanism of spin-2 dark matter which we shall discuss in this section.

So far we have assumed the vacuum configuration, but to discuss a realistic cosmological scenario, we have to add matter components such as radiation and inflaton. 
In general, the graviton mass squared $M^2$ is expected to depend on the matter field through the complicated constraint equations. As a simple example, let us consider a scalar field $\phi$ as a matter field and promote the coupling constants $b_i$ to be functions of $\phi$. In particular $b_0(\phi)$ (or $b_4(\phi)$) is nothing but a potential of the scalar field minimally coupled to the $g$-metric (or the $f$-metric). The theory with $\phi$-dependent coupling constants $b_1,b_2,b_3$ is known as chameleon bigravity~\cite{DeFelice:2017oym, DeFelice:2017gzc} (see also~\cite{Aoki:2020rae} as well as a similar setup in MTMG~\cite{Fujita:2018ehq,Fujita:2019tov}). As $\phi$ evolves in time, the coupling constants $b_i(\phi)$ also change which may realize a phase transition from a Fig.~\ref{fig:node_saddle}-type phase diagram to a Fig.~\ref{fig:saddle_spiral}-type phase diagram. We shall not discuss a concrete realization of this scenario in the present paper because it would be strongly model dependent. However, we have confirmed that there indeed exists a one-parameter change of the coupling constants $b_i(\phi)$ that realizes an adiabatic transition from Fig.~\ref{fig:node_saddle} to Fig.~\ref{fig:saddle_node} and then Fig.~\ref{fig:saddle_spiral}.

In the first stage (Fig.~\ref{fig:node_saddle}), the isotropic universe is unstable due to a tachyonic mass $M_I^2<0$ and a non-zero value of $\beta$ can be spontaneously produced (when the Hubble friction is not too large). Then, $\beta$ eventually reaches the vicinity of the anisotropic fixed point. After the phase transition from Fig.~\ref{fig:node_saddle} to Fig.~\ref{fig:saddle_node}, the anisotropic fixed point turns into unstable one by changing the sign of $M_A^2$ and then $\beta$ starts to deviate from the vicinity of the anisotropic fixed point. As the graviton mass increases (or the Hubble expansion rate $H_g$ decreases), the phase diagram further changes from Fig.~\ref{fig:saddle_node} to Fig.~\ref{fig:saddle_spiral}. As a result, $\beta$ behaves as a dark matter component of the universe around the isotropic fixed point.

In this scenario, the dark matter abundance is roughly estimated by the value of the anisotropic fixed point and the time of phase transition. For simplicity, we assume that the evolution of $\beta$ in the second stage (Fig.~\ref{fig:saddle_node}) is negligible and $M_A$ does not significantly change after the transition. Provided that the phase transition occurs at $H_{g}(a_{\rm tra}) \sim M_{A}(a_{\rm tra}) \sim m$, the present amount of dark matter is computed in the same way as the misalignment mechanism \cite{Marzola:2017lbt, Preskill:1982cy, Abbott:1982af, Dine:1982ah} by replacing the initial amplitude with the fixed-point value. Here we assume the coupling constants $b_i(\phi)$ are of order unity at the transition time and hence $\beta$ and $\xi$ are also approximately of order unity.

\begin{figure}
    \centering
    \includegraphics[width=\linewidth]{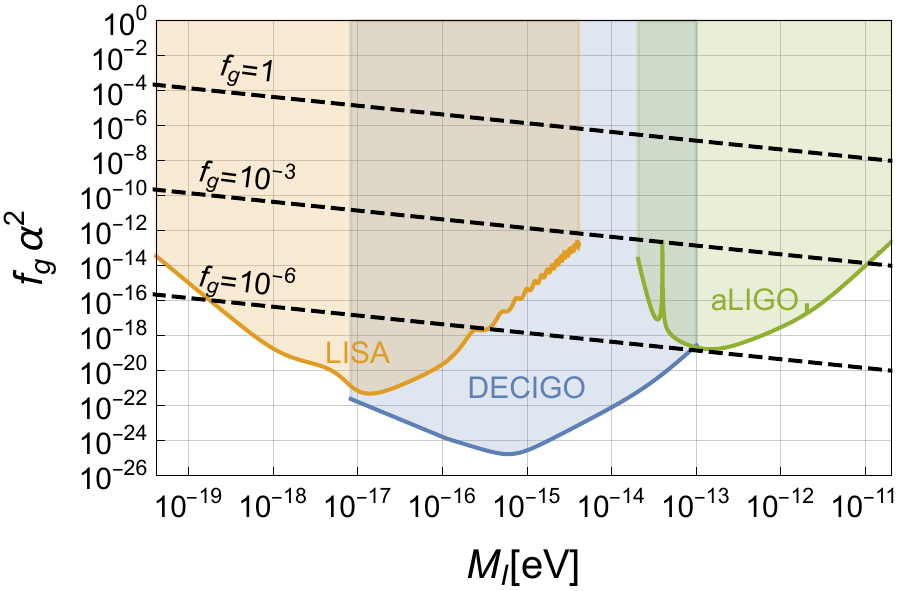}
    \caption{Constrains on spin-2 dark matter from the current and future experiments. 
    The green, blue, and orange region represents the estimate of the detectability of the spin-2 dark matter with $f_g\alpha^2$ by advanced LIGO, DECIGO, and LISA, respectively. In this figure, we use the sensitivity curve in \cite{LIGOsensitivity,Yagi:2011wg,Robson:2018ifk}, and assume 2 years of the observation time (see Appendix \ref{sec:constraint}). 
    The black dashed lines represent the rough estimate of $f_g\alpha^2$ in our production mechanism by using \eqref{eq:fraction}. They are given by fixing the fraction of spin-2 dark matter density to the total dark matter density as $f_g=1,10^{-3},10^{-6}$ in our scenario. The plotted sensitivity of advanced LIGO is consistent with ``optimised sensitivity'' in~\cite{Armaleo:2020efr}.}
    \label{fig:fgalpha2}
\end{figure}

By assuming the transition occurs at the radiation dominant era, the scale factor at the transition time is estimated as
\begin{align}
    a_{\rm tra} \sim \Omega_{r,0}^{1/4} \left(\frac{H_0}{H_g(a_{\rm tra})}\right)^{1/2} 
    \sim \Omega_{r,0}^{1/4} \left(\frac{H_0}{m}\right)^{1/2}\,,
    \label{eq:scalefacotr}
\end{align}
where $\Omega_{r,0}$ is the current density parameter of the radiation components. The energy density of spin-2 dark matter at the transition time can be roughly estimated as $\rho_g(a_{\rm tra})\sim \mpl^2 m_g^2$. Then the current density parameter of spin-2 dark matter is
\begin{align}
    \Omega_{g,0}=\frac{\rho_g(a_{\rm tra})}{\rho_{c,0}}a_{\rm tra}^3
    \sim \frac{\alpha^2}{1+\alpha^2}\Omega_{r,0}^{3/4}\left(\frac{m}{H_0}\right)^{1/2}\,.
\end{align}
This is consistent with the result in \cite{Marzola:2017lbt}. The fraction of the density of spin-2 dark matter to the total dark matter is given by
\begin{align}
    f_g \equiv \frac{\Omega_{g,0}}{\Omega_{\rm DM}} \sim \frac{\alpha^2}{1+\alpha^2} \Omega_{r,0}^{3/4}\left(\frac{m}{H_0}\right)^{1/2}\,.
    \label{eq:fraction}
\end{align}

Since the spin-2 dark matter couples to matter fields in the same way as the massless graviton, a signal caused by the oscillating spin-2 dark matter can be probed by the gravitational wave detectors. As detailed in Appendix \ref{sec:constraint}, the signal depends on the combination $f_g\alpha^2$ and the graviton mass $M_I$. In Fig. \ref{fig:fgalpha2}, we show the detectability of $f_g\alpha^2$ for spin-2 dark matter by advanced LIGO, DECIGO, and LISA.

In our scenario, when the fraction is fixed, we obtain a relation between the graviton mass $M_I\sim m$ and the ratio of the gravitational constants $\alpha$ by using \eqref{eq:fraction}. The values of $f_g \alpha^2$ for several fixed $f_g$ are shown as black dashed lines in Fig. \ref{fig:fgalpha2}. The signal of our scenario is detectable even if the massive spin-2 field only contributes to a small fraction of the total dark matter density in the mass range $m \lesssim 10^{-10} {\rm eV}$.

\section{Conclusion}
\label{sec:discussion}
In the present paper, we have considered the Bianchi type-I solution in the two kinds of bigravity theories without Boulware-Deser ghost, Hassan-Rosen bigravity and Minimal Theory of Bigravity. First, we have identified the background equations for the Bianchi type-I Universe, and found that the background equations are the same in the two theories. Furthermore, we have found fixed points of the background equations with relatively large anisotropy and classified them by local stability. We have also investigated the global stability around the fixed points by showing the phase portraits for all patterns of the local stability.

One of the interesting applications of the anisotropic fixed point is the production of spin-2 dark matter.
The production of spin-2 dark matter corresponds to the production of the initial anisotropy $\beta$ in the Universe. One way to generate the initial amplitude of $\beta$ is a phase transition that changes the stability of anisotropic and isotropic fixed points. The phase transition can be achieved by introducing a matter field. 
Our scenario is somewhat similar to the axion dark matter~\cite{Preskill:1982cy,Abbott:1982af,Dine:1982ah}. In the misalignment mechanism of the axion dark matter, the initial amplitude of axion is generated by a misalignment away from the bottom of the potential in the early universe. 
In our scenario, on the other hand, the ``misalignment'' is spontaneously generated by the instability of the isotropic fixed point even if its initial amplitude is negligibly small, and the size of the ``misalignment'' is fixed when the model is given. 
The rough estimation of the abundance from this production mechanism shows that spin-2 dark matter can account for all or a part of dark matter. 
As shown in Fig.~\ref{fig:fgalpha2}, gravitational wave detectors are expected to be able to search ultralight spin-2 dark matter in a certain range of the graviton mass even if its fraction to all dark matter is small.

\begin{acknowledgements}
    We would like to thank Hiroki Takeda for insightful comments. The work of Y.M. was supported by the establishment of university fellowships towards the creation of science technology innovation.
    This work was supported in part by Japan Society for the Promotion of Science (JSPS) Grants-in-Aid for Scientific Research No.~20K14468 (K.A.), No.~18K13537 (T.F.), No.~20H05854 (T.F.) and No.~17H02890 (S.M.), No.~17H06359 (S.M), and by World Premier International Research Center Initiative, MEXT, Japan. 
\end{acknowledgements}

\appendix

\section{Hamiltonian formulation of Minimal Theory of Bigravity}
\label{sec:hamilton_formalism}
In this section, we derive the background equation in Bianchi type-I Universe for the Minimal theory of Bigravity through the Hamilton formalism. The Minimal Theory of Bigravity is originally constructed with Hamiltonian to impose an appropriate constraint, and thus it looks relatively simple in the Hamilton formalism. We define the canonical momentum associated with $a_g,a_f,\beta_g,\beta_f$ as
\begin{align}
    P_{g} &= \frac{\partial L}{\partial \dot{a}_g}\,,
    \quad
    P_{f} = \frac{\partial L}{\partial \dot{a}_f}\,,  
    \\
    Q_{g} &= \frac{\partial L}{\partial \dot{\beta}_g}\,,  
    \quad
    Q_{f} = \frac{\partial L}{\partial \dot{\beta}_f}\,. 
\end{align}
The mini-superspace Hamiltonian in Bianchi type-I Universe is obtained by Legendre transformation of the Lagrangian in \eqref{eq:action_const} as
\begin{align}
    H &= P_{g}\dot{a}_g+P_{f}\dot{a}_f+Q_{g}\dot{\beta}_g+Q_{f}\dot{\beta}_f-L
    \nn
    &=\calC_{N_g}N_g+\calC_{N_f}N_f+\calC_{\lambda}\lambda\,,
\end{align}
where
\begin{widetext}
\begin{align}
    \calC_{N_g}&=-\frac{m^4\mpl^2a_g^3\lambda^2}{8N_g^2}\big[-3b_3^2-4b_2b_3(e^{-2\beta}+2e^{\beta})\xi-2(2b_2^2+b_1b_3)(2e^{-\beta}+e^{2\beta})\xi^2-12b_1b_2\xi^3+b_1^2(e^{4\beta}-4e^{\beta})\xi^4\big]
    \nn
    &+\frac{m^2\mpl^2a_g^3}{2}\big[b_4+b_3(e^{-2\beta}+2e^{\beta})\xi+b_2(2e^{-\beta}+e^{2\beta})\xi^2+b_1\xi^3\big]+\frac{-a_g^2 P_g^2+Q_g^2}{12\mpl^2 a_g^3}\,,
    \\
    \calC_{N_f}&=\frac{m^4\mpl^2a_g^3\lambda^2}{8\alpha^2N_f^2\xi}\big[b_3^2(-e^{-\beta}+4e^{-\beta})+12b_2b_3\xi+2(2b_2^2+b_1b_3)(e^{-2\beta}+2e^{\beta})\xi^2+4b_1b_2(2e^{-\beta}+e^{2\beta})\xi^3+3b_1^2\xi^4\big]
    \nn
    &\frac{m^2\mpl^2a_g^3}{2}\big[b_3+b_2(e^{-\beta}+2e^{\beta})\xi+b_1(2e^{-\beta}+e^{2\beta})\xi^2+b_0\xi^3\big]+\frac{-a_g^2\xi^2 P_f^2+Q_f^2}{12\alpha^2 \mpl^2 a_g^3 \xi^3}\,,
    \\
    \calC_\lambda&=-\frac{m^4\mpl^2a_g^3\lambda}{4\alpha^2N_gN_f\xi}(b_3+2b_2e^\beta\xi+b_1e^{2\beta}\xi^2)\big\{N_g[b_3(-e^{-4\beta}+4e^{-\beta})+2b_2(e^{-3\beta}+2)\xi+3b_1e^{-2\beta}\xi^2]
    \nn
    &-\alpha^2N_f\xi[-3b_3-2b_2(2e^{-2\beta}+e^{\beta})\xi+b_1(-4e^{-\beta}+e^{2\beta})\xi^2]\big\}
    \nn
    &+\frac{m^2a_g}{12\alpha^2\xi}\big\{\big[b_3e^{-2\beta}+2b_3e^{\beta}+2b_2(2e^{-\beta}+e^{2\beta})\xi+3b_1\xi^2\big]P_f-\alpha^2\xi\big[3b_3+2b_2(e^{-2\beta}+2e^{\beta})\xi+b_1(2e^{-\beta}+e^{2\beta})\xi^2\big]P_g\big\}
    \nn
    &+\frac{m^2(e^{\beta}-e^{-2\beta})}{6\alpha^2\xi^2}\big[(b_3+b_2e^{\beta}\xi)Q_f+\alpha^2\xi^3(b_2+b_1e^{\beta}\xi)Q_g\big]\,.
    \label{eq:constraint_lambda}
\end{align}
\end{widetext}
Then we immediately get the constraint equations $\calC_{N_g}\approx 0,\calC_{N_f}\approx 0,\calC_{\lambda}\approx 0$. We can also obtain the canonical equations illustrated by
 \begin{align}
     \dot{P}_g &= -\frac{\partial H}{\partial a_g}\,,
     &\dot{P}_f &= -\frac{\partial H}{\partial a_f}\,,
     \\
     \dot{a}_g &= \frac{\partial H}{\partial P_g}\,,
     &\dot{a}_f &= \frac{\partial H}{\partial P_f}\,,
    \label{eq:canonical_scale_factor}
\end{align}
\begin{align}
     \dot{Q}_g &= -\frac{\partial H}{\partial \beta_g}\,,
     &\dot{Q}_f &= -\frac{\partial H}{\partial \beta_f}\,,
     \\
     \dot{\beta}_g &= \frac{\partial H}{\partial Q_g}\,,
     &\dot{\beta}_f &= \frac{\partial H}{\partial Q_f}\,.
\label{eq:canonical_beta}
 \end{align}
The consistency of the constraint equation requires that time derivatives of the constraint equations have to vanish. Substituting \eqref{eq:constraint_lambda}, \eqref{eq:canonical_scale_factor}, and \eqref{eq:canonical_beta} into $\dot{\calC}_{N_g} \approx 0$, we obtain
\begin{align}
    \lambda F_1\left[\lambda,a_g,a_f,\beta_g,\beta_f,P_g,P_f,Q_g,Q_f\right] = 0\,.
\end{align}
The function is linear in $\lambda$, then we get two branches of the solution, $\lambda\approx 0$ and $F_1 \approx 0$. Similarly, $\dot{\calC}_{N_g} \approx 0$ gives
\begin{align}
    \lambda F_2\left[\lambda,a_g,a_f,\beta_g,\beta_f,P_g,P_f,Q_g,Q_f\right]=0\,.
\end{align}
Then it in principle gives two branches of the solution, $\lambda \approx 0$ and $F_1\approx0\land F_2\approx0$. Although MTBG is intended to give the same background equations as HRBG in the homogeneous Universe, $F_1\approx0\land F_2\approx0$ leads to an additional constraint to the background. Furthermore, it can be shown that the background solution with $F_1\approx0\land F_2\approx0$ does not work well at least for the isotropic Universe, thus, we select $\lambda\approx 0$.
Since the differences in the equations from Hassan-Rosen bigravity are terms with $\lambda$, we now confirm that the equations of the Minimal Theory of Bigravity are identical with those of the Hassan-Rosen bigravity.

\section{Probing spin-2 dark matter with gravitational wave detectors}

\label{sec:constraint}
In this section, we briefly show the detectability of spin-2 dark matter. The main result is shown in Fig.\ref{fig:fgalpha2}. 
Our analysis is similar to that in Ref.~\cite{Armaleo:2020efr}.

\subsection{Perturbations around the Minkowski spacetime}

We will consider the action of bigravity with matter field $\psi_{\rm m}$ which couples only to the $g$-metric:
\begin{align}
    S = S_g+S_{\rm m}[\psi_{\rm m},g_{\mu\nu}]\,,
\end{align}
where $S_g$ is defined by \eqref{eq:bigravity_action}. In order to analyze the responses of the gravitational wave detector, we define the metric perturbations around the Minkowski spacetime by
\begin{align}
    \delta g_{\mu \nu}&:=g_{\mu \nu}-\eta_{\mu \nu}\,, 
    \\
    \delta f_{\mu \nu}&:=f_{\mu \nu}-\eta_{\mu \nu} .
\end{align}
Note that either $\delta g_{\mu\nu}$ or $\delta f_{\mu\nu}$ is not a mass eigenstate. At the linear order, the mass eigenstate is given by
\begin{align}
h_{\mu \nu}&:=\frac{\kappa_{f}}{\kappa_{g} \kappa} \delta g_{\mu \nu}+\frac{\kappa_{g}}{\kappa_{f} \kappa} \delta f_{\mu \nu} \\
\varphi_{\mu \nu}&:=\frac{1}{\kappa}\left(\delta g_{\mu \nu}-\delta f_{\mu \nu}\right)\,.
\end{align}
The quadratic-order action is then
\begin{align}
    S_{2} &= \int d^4x \Bigg[\calL_{\rm EH}[h]+\calL_{\rm EH}[\varphi]+\calL_{\rm FP}[\varphi]
    \nn
    &+\frac{1}{2\mpl}h_{\mu\nu}T_{\rm m}^{\mu\nu}+\frac{1}{2M_{G}}\varphi_{\mu\nu}T_{\rm m}^{\mu\nu}\Bigg]\,,
\end{align}
where
\begin{align}
    M_{\mathrm{pl}}:=\frac{\kappa}{\kappa_{g} \kappa_{f}}, \quad M_{G}:=\frac{\kappa}{\kappa_{g}^{2}}=\frac{\kappa_{f}}{\kappa_{g}} M_{\mathrm{Pl}}\,,
\end{align}
and for an arbitrary $\chi_{\mu\nu}$, we define
\begin{align}
   \calL_{\rm EH}[\chi] &:= \frac{1}{8}\big[(2\p_\nu \chi_{\mu\rho}-\p_\rho \chi_{\mu\nu})\p^\rho \chi^{\mu\nu}
   \nn
   &+(\p^\mu \chi-2\p_\nu \chi^{\mu\nu})\p_\mu \chi\big]\,,
   \\
   \calL_{\rm FP}[\chi] &:= \frac{M^2}{8} (\chi^2-\chi^{\mu\nu}\chi_{\mu\nu})\,,
\end{align}
with the mass of spin-2 dark matter $M$, and we have used the notation $\chi=\chi^\mu{}_\mu$. Ultralight spin-2 dark matter in our Galaxy is modeled by
\begin{align}
    \varphi_{ij}=\sum_\lambda \varphi_{0,\lambda} e^\lambda_{ij} \cos (\omega t-\bm{k}\cdot\bm{x}+\delta_{\tau}(t))\,,
\end{align}
where $\delta_\tau(t)$ is a time-dependent phase factor, which evolves on 
the coherent timescale $\tau=2\pi/(Mv_{\rm DM}^2)$. 
Since the typical dark matter velocity in our Galaxy is $v_\mathrm{DM}\sim 10^{-3}$, we can use the non-relativistic approximation $\omega \sim M$. In this model, the dark matter density is given by
\begin{align}
  \rho_g=\frac{1}{4}\left<{\dot{\varphi}_{ij}\dot{\varphi}}_{ij}\right> \simeq \frac{M^2}{4}\sum_\lambda \left<\varphi_{0,\lambda}^2\right>\,.
\end{align}
where the symbol $\braket{\cdots}$ denotes the spacetime average. We have used the fact $\braket{\cos^2 (Mt)} = 1/2$ and $e_{ij}^\lambda e_{ij}^{\lambda'}=2\delta^{\lambda\lambda'}$. In the following, we assume massive graviton with only helicity two modes:
\begin{align}
    \varphi_{0} &:= \sqrt{\braket{\varphi_{0,+}^2}} = \sqrt{\braket{\varphi_{0,\times}^2}} = \frac{\sqrt{2\rho_g}}{M}\,, 
    \\
    \sqrt{\braket{\varphi_{0,x}^2}} &= \sqrt{\braket{\varphi_{0,y}^2}} = \sqrt{\braket{\varphi_{0,b}^2}} =0\,.
\end{align}

\subsection{Signal in a gravitational-wave detector}

The $g$-metric, which is coupled to the matter fields, is given by
\begin{align}
    g_{\mu\nu} = \eta_{\mu\nu} + \frac{h_{\mu\nu}}{\mpl} + \frac{\varphi_{\mu\nu}}{M_G}\,.
\end{align}
The signal for the gravitational wave detector from the massive graviton is given by operating the detector tensor $D^{ij}=(\hat{x}^i\hat{x}^j-\hat{y}^i\hat{y}^j)/2$ to the fluctuation,
\begin{align}
    h(t) &= \frac{1}{M_G} D^{ij}\varphi_{ij} 
    \nn
    &= \frac{\alpha\varphi_0}{\mpl}  [F_+(\theta, \phi, \psi) + F_\times(\theta, \phi, \psi)]
    \nn
    &\times \cos (\omega t-\bm{k}\cdot\bm{x}+\delta_{\tau}(t))
    \,.
\end{align}
where $F_{+},F_{\times},\cdots$ are antenna pattern functions which depend on the sky location $(\theta,\phi)$ and polarization angle $\psi$. For advanced LIGO, the antenna pattern functions are given by
\begin{align}
  F_{+}(\theta, \phi, \psi)&=\frac{1}{2}(1+\cos^{2}\theta)\cos(2\phi)\cos(2\psi)\nn
  &-\cos\theta\sin(2\phi)\sin(2\psi)\,,
  \\
  F_{\times}(\theta, \phi, \psi)
  &=\frac{1}{2}(1+\cos^{2}\theta)\cos(2\phi)\sin(2\psi)
  \nn
  &+\cos\theta\sin(2\phi)\cos(2\psi)\,.
\end{align}
The sky/polarization average of squared antenna pattern functions are given by
\begin{align}
  \mathcal{R} = \left< F_{+}^{2}\right>=\left< F_{\times}^{2}\right>=\frac{1}{5}\,, \quad \left< F_{+} F_{\times}\right>=0\,,
\end{align}
where the bracket $\left<\cdots\right>$ denotes
\begin{align}
   \left<\cdots\right>=\frac{1}{4\pi^{2}}\int_{0}^{\pi}d\psi\int_{0}^{2\pi}d\phi\int_{0}^{\pi}d\theta\sin\theta(\cdots)\,.
\end{align}

For LISA, the antenna pattern functions depend on the frequency, and their sky/polarization average $\mathcal{R}$ is given by \cite{Robson:2018ifk}
\begin{align}
\mathcal{R} = \frac{3}{10} - \frac{507}{5040} \left( \frac{f}{f_{*}}\right)+ \cdots
\end{align}
where $f_{*}=19.09$ mHz is the peak frequency.

The threshold of the detection signal can be estimated by
\begin{align}
    \left<h^2\right> = \frac{S_n(\frac{M}{2\pi})}{T_{\rm eff}}\,,
\end{align}
where $S_n$ is the one-sided noise spectrum of each detector, and $T_{\rm eff}$ is the effective observation time 
that takes into account the coherent timescale $\tau$~\cite{Budker:2013hfa}
\begin{align}
    T_{\rm eff}=\left\{
    \begin{array}{cc}
        T_{\rm obs} & (T_{\rm obs}<\tau)  \\
        \sqrt{\tau T_{\rm obs}} & (T_{\rm obs} \geq \tau)\,
    \end{array}
    \right..
\end{align}
Here, the time-averaged signal is
\begin{align}
    \left<h^2\right> = \frac{2\alpha^2f_g\rho_{\rm DM}}{5 \mpl^2 M^2}\,,
\end{align}
where $\rho_{\rm DM}\simeq 0.3~{\rm GeV/cm^3}$ is the local dark matter density, and $f_g=\rho_g/\rho_{\rm DM}$ is the spin-2 dark matter fraction of the total dark matter density. 
Plugging $T_{\rm obs}=2$ years and the noise spectra given in Ref.~\cite{LIGOsensitivity,Yagi:2011wg,Robson:2018ifk}, we obtain the sensitivity curves for $\alpha^2 f_g$ shown in Fig.~\ref{fig:fgalpha2}.

\bibliography{main}

\end{document}